\documentstyle[a4,11pt]{article}

\date{}

\begin{document}

\title{\bf Cosmological perturbations on a magnetised Bianchi~I
background}

\author{Christos G. Tsagas\thanks{e-mail address: christos.tsagas@port.ac.uk}
and Roy Maartens\thanks{e-mail address: roy.maartens@port.ac.uk}\\
{\small Relativity and Cosmology Group, Division of Mathematics
and Statistics,}\\ {\small Portsmouth University,
Portsmouth~PO1~2EG, England}}

\maketitle

\begin{abstract}
Motivated by the isotropy of the CMB spectrum, all existing
studies of magnetised cosmological perturbations employ FRW
backgrounds. However, it is important, to know the limits of this
approximation and the effects one loses by neglecting the
anisotropy of the background magnetic field. We develop a new
treatment, which fully incorporates the anisotropic magnetic
effects by allowing for a Bianchi~I background universe. The
anisotropy of the unperturbed model facilitates the closer study
of the coupling between magnetism and geometry. The latter leads
to a curvature stress, which accelerates positively curved
perturbed regions and balances the effect of magnetic pressure
gradients on matter condensations. We argue that the tension
carried along the magnetic force-lines is the reason behind these
magneto-curvature effects. For a relatively weak field, we also
compare to the results of the almost-FRW approach. We find that
some of the effects identified by the FRW treatment are in fact
direction dependent, where the key direction is that of the
background magnetic field vector. Nevertheless, the FRW-based
approach to magnetised cosmological perturbations remains an
accurate approximation, particularly on large scales, when one
looks at the lowest order magnetic impact on gravitational
collapse. On small scales however, the accuracy of the perturbed
Friedmann framework may be compromised by extra shear effects.
\end{abstract}

%%%%%%%%%%%%%%%%%%%%%%
\section{Introduction}
%%%%%%%%%%%%%%%%%%%%%%
Over the years, the implications of primordial magnetic fields for
the formation and the evolution of the observed structure have
been the subject of continuous theoretical investigation. The
majority of theses studies are Newtonian, or semi-relativistic,
and all of them are confined within perturbed
Friedmann-Robertson-Walker (FRW) models. Mathematically speaking,
however, the spatial isotropy of the Friedmann universe is not
compatible with the presence of large-scale magnetic fields.
Nevertheless, one may still preserve the FRW symmetries by
postulating a random background field (e.g. \cite{W,SB}), or by
allowing for a uniform field that is too weak to destroy the
Robertson-Walker isotropy (e.g. \cite{RR}-\cite{TB1}). In either
approach the anisotropy induced by the magnetic field to the
zero-order stress tensor of the universe is treated as a
perturbation. Both treatments, apart from their different initial
set up, proceed essentially in the same way and they should reach
the same conclusions.

Strong motivation for adopting an FRW background comes from
observation. Current measurements of the Cosmic Microwave
Background (CMB) anisotropy suggest an energy density for the
primeval field much smaller than that of matter at recombination
\cite{BFS}. Similar limits, though less stringent, are placed at
nucleosynthesis by Helium-4 measurements \cite{GR}. One may also
use dissipative arguments to limit the strength of primordial
magnetic fields on small scales \cite{JKO}. Intuitively, the
impact of such a weak field on the isotropy of the smooth FRW
spacetime should be negligible. Nevertheless, pending on more
detailed treatments, one might still ask:
\begin{itemize}
\item
What is the exact nature of the effects induced by the anisotropic
magnetic stresses?
\item
How important are the deviations from the results based on
perturbed FRW models?
\end{itemize}
We address these questions by employing a Bianchi~I background,
the simplest anisotropic cosmology that can naturally accommodate
large-scale magnetic fields. We also use the covariant Lagrangian
perturbation formalism \cite{EB}, which has proved powerful in
dealing with issues of similar complexity in the past. This
article extends the work of \cite{TB1}, where the exact equations
of a general magnetised cosmology were derived and subsequently
linearised about an FRW spacetime. Here, we linearise the same
equations about a Bianchi~I universe supplied with a large-scale
magnetic field and a perfectly conducting ideal fluid. Our main
aim is to identify and analyse the kinematic and dynamic effects
of the anisotropic magnetic stresses. In the process we will also
test the accuracy of the FRW-based treatments of magnetised
cosmological perturbations.

We provide the full set of the linear equations and then isolate
the effects of the background magnetic stresses, by treating the
field as an eigenvector of the shear tensor to zero order. This
way we can identify the deviations from the almost-FRW results
that are caused purely by the background magnetic stresses. Note
that background anisotropies that are not related to the magnetic
presence are not relevant to our study and therefore not included.
At the linear level, however, the field is not assumed to be a
shear eigenvector. Thus, in the perturbed universe the shear is no
longer the pure result of the magnetic stresses. We consider a
universe filled with a single perfectly conducting barotropic
fluid and compare to the predictions of the FRW-based treatments.
Generally speaking, the impact of a cosmological magnetic field
has an isotropic as well as an anisotropic component. The former
results from the magnetic contributions to the total energy
density and isotropic pressure, whereas the latter is due to the
magnetic stresses. Our analysis shows that although the
Friedmannian limit identifies the isotropic component exactly, it
can only address the `average' anisotropic effects of the field.
Note that the `average' is taken over the three linearly
independent spatial directions rather than over the whole space.
Thus, as one should expect, the isotropy of the background
spacetime sets certain limits to the accuracy of the standard
treatment. In this report we show that some of the magnetic
effects identified by the FRW approximation are in fact direction
dependent, with the key direction determined by the zero-order
field vector. Even when the field is relatively weak, certain
quantities, such as the curvature, continue to contribute
differently parallel to the magnetic force-lines than orthogonal
to them. Clearly, in the linear regime, such directional
dependence results from the background anisotropy and it is not
accessible through an FRW background.

Of particular interest are the magnetohydrodymanical effects on
the average volume expansion of the universe. We find that density
perturbations co-linear to the background field contribute more
efficiently to the deceleration rate. Also, only spatial curvature
perturbations parallel to the zero-order magnetic vector affect
the volume expansion. The latter effect is of geometrical origin,
resulting from the vectorial nature of the magnetic field. It was
first identified in \cite{TB2}, although there the isotropy of the
FRW background meant that the effect was independent of direction.
Here, as in \cite{TB2}, the magneto-curvature term in the
Raychaudhuri equation decelerates negatively curved regions, but
accelerates those with positive local spatial curvature. The
crucial difference made by the background anisotropy, is that it
reveals the prominent role played by the field lines themeselves.
We can now attribute the unconventional magneto-geometrical
effects to the negative pressure, the tension, experienced along
the magnetic lines of force. This property of the field, together
with the coupling between magnetism and geometry, gives rice to a
{\em curvature stress}. The latter reacts to the effect of local
magnetic pressure gradients by modifying the expansion rate of the
region accordingly. This relativistic effect is analogous to the
classical curvature stress exerted by field lines with a nonzero
local radius of curvature.

Direction dependent contributions are also identified in the
evolution equations of the density gradients. Curvature effects,
for example, are again confined in the direction of the background
field. At the weak field limit and on super-horizon scales, this
change in the curvature contribution is what separates the
Bianchi~I perturbation equations from their FRW counterparts.
Nevertheless, such directional effects can only alter the lowest
order magnetic impact during the radiation era. Even then, the
predictions of the FRW approximation still hold, at least in
certain critical cases. Our results verify the point made in
\cite{TM}, that magneto-curvature effects tend to counterbalance
the `pure' magnetic impact on the evolution of density gradients.
In particular, for zero curvature input the field was found to
suppress slightly the growth of the density contrast \cite{TB2},
in agreement with the Newtonian treatment of \cite{RR}. However,
when curvature contributions were included in the equations, this
inhibiting effect vanished \cite{TM}. Here, we verify these
statements and argue that the magneto-geometric contributions can
also reverse the pure magnetic impact on gravitational collapse.
When the curvature perturbation along the field lines is maximum,
we find that the density contrast grows faster than in standard
non-magnetised cosmological models. Thus, the combined
magneto-curvature action opposes the effect of the pure field on
local matter condensations. Again, the explanation comes from the
tension experienced along force-lines of the field. When coupled
to geometry, this magnetic property leads to a curvature stress,
which always tries to balance the effect of the magnetic pressure
gradients.

On sub-horizon scales, we find that curvature complexities are
accompanied by similar direction dependent shear effects, even
when the field is weak. These shear contributions result from the
kinematical implications of the background magnetic stresses, and
do not appear in an almost-FRW treatment. One might be able to
invoke fluid `backreaction' arguments and ignore, to first order,
the aforementioned small-scale shear effects. In this case, one
recovers the standard FRW linear equations, provided that the
curvature perturbation is averaged. Nevertheless, one should be
rather careful when discussing the small-scale magnetic effects
within perturbed FRW cosmologies. In contrast, super-horizon
weakly magnetised density perturbations proceed basically as
predicted by the Friedmann approximation.

This paper is organised as follows: Section 2 provides a brief
review of the covariant and gauge-invariant approach to cosmology
with references for further reading. In section 3 we linearise the
exact equations about a magnetised Bianchi~I background.
Particular properties of the unperturbed model are discussed in
section 4. In section 5 we apply the linear equations to the case
of a barotropic medium, and examine the field impact on the
model's kinematic and dynamic evolution. Section 6 addresses the
weak field limit and section 7 discusses solutions of the
perturbation equations. Finally, we summarise our conclusions in
section 8.

%%%%%%$%%%%%%%%%%%%%%%%%%%%%%%%%%
\section{The Covariant Formalism}
%%%%%%%%%%%%%%%%%%%%%%%%%%%%%%%%%
In the covariant Lagrangian approach to cosmology (see \cite{EvE}
for an up-to-date review) one makes a physical choice for the
4-velocity $u_a$ of the matter. Then, one uses $h_{ab}\equiv
g_{ab}-u_au_b$, where $g_{ab}$ is the spacetime metric, to project
into the instantaneous rest spaces of the comoving
observers.\footnote{We employ geometrised units with $c=1=8\pi G$
and our metric signature is $(-\,+\,+\,+)$.} Note that these rest
spaces form an integrable 3-dimensional hypersurface only for zero
vorticity. Nevertheless, we will keep referring to them as
`spatial' or `3-spaces' for simplicity. Thus, the fluid 4-velocity
introduces a unique 1+3 splitting of the spacetime into time and
space. We can now decompose tensor fields and tensor equations
into their timelike and spacelike parts. Without going into much
detail, we provide some key decompositions essential to our
analysis.

%%%%%%%%%%%%%%%%%%%%%%%%%%%%%%%%
\subsection{The Basic Equations}
%%%%%%%%%%%%%%%%%%%%%%%%%%%%%%%%
The kinematical properties of the motion arise from splitting the
covariant derivative of $u_a$ into its irreducible parts
\begin{equation}
\nabla_bu_a={\textstyle{1\over3}}\Theta h_{ab}+
\sigma_{ab}+\varepsilon_{abc}\omega^c-\dot{u}_au_b\,, \label{ui;j}
\end{equation}
where $\varepsilon_{abc}$ is the rest-space volume element,
$\Theta\equiv \tilde{\nabla}^au_a$ is the rate of the volume
expansion, $\sigma_{ab}\equiv\tilde{\nabla}_{\langle
a}u_{b\rangle}$ is the shear,
$\omega_a\equiv-{\textstyle{1\over2}}{\rm curl}u_a$ is the
vorticity and $\dot{u}_a\equiv u^b\tilde{\nabla}_bu_a$ is the
acceleration.\footnote{Following \cite{M1}, we use angled brackets
for the projected, symmetric, trace-free (PSTF) part of tensors
and for the orthogonal projections of vectors. The same notation
also represents orthogonally projected covariant time
derivatives.} The volume expansion defines an average length scale
$S$ via $\dot{S}/S={\textstyle{1\over3}}\Theta$.\footnote{For
comparison reasons, the current article adopts the notation of
\cite{TB1} with a number of changes motivated by \cite{EvE} and
\cite{M1}. In particular, totally projected covariant derivatives
are now represented by $\tilde{\nabla}$ instead of
$\mbox{}^{(3)}\nabla$, $\Pi_{ab}$ has replaced $M_{ab}$ as the
anisotropic magnetic pressure tensor, the 3-Ricci tensor and the
3-Ricci scalar have changed from $\mbox{}^{(3)}R_{ab}$ and $K$ to
${\cal R}_{ab}$ and ${\cal R}$ respectively, while the scalar
${\cal B}$ represents now the dimensionless ratio
$S^2\tilde{\nabla}^2H^2/H^2$ rather than the Laplacian
$\tilde{\nabla}^2H^2$. Other changes include a difference in the
definition of the Alfv\'en speed (see \S 4.2.2) and the use of the
vorticity vector $\omega_a$ instead of the antisymmetric tensor
$\omega_{ab}$. Finally, we employ the `streamline' definitions for
the covariant curls of spacelike vectors and tensors (see
\cite{M1}).}

The medium is characterised by its energy-momentum tensor.
Relative to a comoving observer, the stress tensor of an imperfect
fluid decomposes as
\begin{equation}
T^*_{ab}=\mu^*u_au_b+p^*h_{ab}+ 2q^*_{(a}u_{b)}+\pi^*_{ab}\,,
\label{fTij}
\end{equation}
where $\mu^*$ and $p^*$ are respectively the relativistic energy
density and the isotropic pressure, $q^*_a$ is the energy flux
vector and $\pi^*_{ab}$ is the PSTF tensor representing
anisotropic pressures. The detailed physics is encoded in the
equations of state. For a perfect fluid, with zero flux and
anisotropic stresses, Eq. (\ref{fTij}) reduces to
\begin{equation}
T_{ab}=\mu u_au_b+ph_{ab}\,.  \label{pfTij}
\end{equation}

Similarly, in the absence of electric fields, the stress tensor of
a pure magnetic field $H_a$ splits as
\begin{equation}
T_{ab}={\textstyle{1\over2}}H^2u_au_b+
{\textstyle{1\over6}}H^2h_{ab}+\Pi_{ab}\,,  \label{mTij}
\end{equation}
where $H^2\equiv H_aH^a$. The latter provides a measure of the
magnetic energy density and isotropic pressure, while
\begin{equation}
\Pi_{ab}\equiv{\textstyle{1\over3}}H^2h_{ab}-H_aH_b
\equiv-H_{\langle a}H_{b\rangle}  \label{Pi}
\end{equation}
is the PSTF tensor that represents the anisotropic magnetic
stresses.

The Weyl tensor $C_{abcd}$ represents the long-range gravitational
field, namely tidal forces and gravity waves. Together with the
Ricci tensor $R_{ab}$, which is decided locally by matter, it
determines the Riemann curvature tensor $R_{abcd}$ completely. For
a comoving observer we have \cite{MB}
\begin{equation}
C_{ab}{}{}^{cd}=4\left(u_{[a}u^{[c}+h_{[a}{}^{[c}\right)E_{b]}{}^{d]}
+2\varepsilon_{abe}u^{[c}H^{d]e}+2u_{[a}H_{b]e}\varepsilon^{cde}\,,
\label{Wt}
\end{equation}
where $E_{ab}=C_{acbd}u^cu^d$ and
$H_{ab}={\textstyle{1\over2}}\varepsilon_{acd}C^{cd}{}{}_{be}u^e$
are respectively the `electric' and `magnetic' parts of
$C_{abcd}$.

%%%%%%%%%%%%%%%%%%%%%%%%%%%%%%%%%%%%%%%%%%%%%%%%
\subsection{The Basic Gauge-invariant Variables}
%%%%%%%%%%%%%%%%%%%%%%%%%%%%%%%%%%%%%%%%%%%%%%%%
The covariant Lagrangian formalism is the foundation for a
covariant and gauge-invariant perturbation theory \cite{EB}, which
provides an alternative to the metric-based gauge-invariant
formalism \cite{Ba}. In the covariant approach, one deals with the
inhomogeneous and anisotropic spacetime directly, instead of
perturbing away from the smooth background universe. Thus, one
obtains the exact (fully non-linear) equations first, before
linearising them about a chosen background. This is clearly a
major advantage, because it allows one to address cosmological
models more general than the perturbed FRW universes. Indeed, the
covariant approach has already been employed to study perturbed
Bianchi~I cosmologies filled with a non-magnetised perfect fluid
\cite{Du}. Here, we use the same formalism to analyse, for the
first time, almost Bianchi~I spacetimes permeated by a large-scale
magnetic field.

The covariant approach utilises the Stewart-Walker lemma
\cite{SW}, to define a set of gauge-independent tensors. These
describe inhomogeneities in the key quantities and have a
transparent geometrical and physical interpretation. For our
purposes, the basic gauge-invariant variables are \cite{TB1}
\begin{equation}
{\cal D}_a\equiv\frac{1}{\mu}S\tilde{\nabla}_a\mu\,,\hspace{10mm}
{\cal Z}_a\equiv S\tilde{\nabla}_a\Theta \hspace{10mm}\mbox{\rm
and}\hspace{10mm} {\cal M}_{ab}\equiv S\tilde{\nabla}_bH_a\,.
\label{giv}
\end{equation}
The above respectively represent spacelike variations in the
energy density of the fluid, the volume expansion and the magnetic
field vector, as these are seen by a pair of neighbouring
fundamental observers.\footnote{in a perturbed Bianchi~I spacetime
the aforementioned geometrical interpretation of the key
inhomogeneity variables holds only for weak shear (see Appendix).}
They vanish in an exact Bianchi~I spacetime, thus satisfying the
Stewart-Walker criterion for gauge-invariance. Note that gradients
in the fluid pressure are represented by the gauge-independent
variable $Y_a\equiv \tilde{\nabla}_ap$. For a barotropic fluid,
however, $Y_a$ is always given in terms of ${\cal D}_a$.

%%%%%%%%%%%%%%%%%%%%%%%%%%%%%%%%%%%%%%%%%%%%%%%%%%
\section{Perturbed Magnetised Bianchi~I Universes}
%%%%%%%%%%%%%%%%%%%%%%%%%%%%%%%%%%%%%%%%%%%%%%%%%%
The non-linear equations that describe density perturbations of a
perfectly conducting medium in the presence of a large-scale
magnetic field have been derived in \cite{TB1}. Here, we linearise
these equations about a Bianchi~I spacetime. Covariantly, the
dynamics of the model is characterised by the energy density
($\mu$) and the pressure ($p$) of the matter. The magnetic
presence also brings into play the field's energy density and
isotropic pressure (both expressed via $H^2$), as well as
anisotropic stresses represented by $\Pi_{ab}$. Kinematically, the
Bianchi~I cosmology is determined by the average expansion scalar
($\Theta$) and by the shear tensor ($\sigma_{ab}$), which
describes departures from the isotropic expansion. Anisotropies in
the spacetime curvature propagate via the electric Weyl tensor
($E_{ab}$), while the spatial sections of the Bianchi~I spacetime
are flat. During linearisation all these variables are treated as
zero-order. First-order quantities are those that vanish in the
background and therefore satisfy the requirement for gauge
invariance \cite{SW}. These are the acceleration ($\dot{u}_a$),
the vorticity ($\omega_a$), all variables representing spatial
curvature, the magnetic Weyl tensor ($H_{ab}$) and the
inhomogeneity variables ${\cal D}_a$, $Y_a$, ${\cal Z}_a$ and
${\cal M}_{ab}$. Terms of higher perturbative order will be
disregarded.

%%%%%%%%%%%%%%%%%%%%%%%
\subsection{The Medium}
%%%%%%%%%%%%%%%%%%%%%%%
According to Eqs. (\ref{pfTij}) and (\ref{mTij}), the
energy-momentum tensor for a magnetised spacetime filled with a
single perfectly conducting ideal fluid is
\begin{equation}
T_{ab}=\left(\mu+{\textstyle{1\over2}}H^2\right)u_au_b+
\left(p+{\textstyle{1\over6}}H^2\right)h_{ab}+ \Pi_{ab}\,.
\label{Tij}
\end{equation}
where the assumption of infinite conductivity ensures the absence
of electric fields \cite{TB1}. Comparing Eqs. (\ref{Tij}) and
(\ref{fTij}) we see that our magnetised medium corresponds to an
imperfect fluid with $ \mu^*=\mu+{\textstyle{1\over2}}H^2$,
$p^*=p+{\textstyle{1\over6}}H^2$, $q_a^*=0$ and $
\pi^*_{ab}=\Pi_{ab}\equiv-H_{\langle a}H_{b\rangle}$. Thus,
magnetism contributes to the total energy density and isotropic
pressure, while it is solely responsible for the anisotropic
stresses. Note that zero electric field implies a vanishing
Poynting vector, which in turn explains the absence of a energy
flux vector in Eq. (\ref{Tij}).

%%%%%%%%%%%%%%%%%%%%%%%%%%%%%%%%%
\subsection{The Linear Equations}
%%%%%%%%%%%%%%%%%%%%%%%%%%%%%%%%%
Next, we provide the key relations that describe the linear
evolution of our magnetised Bianchi~I cosmology. Unless stated
otherwise, all equations derive from the non-linear formulae given
in \cite{TB1}.\footnote{For a quick cross-check, the reader may
compare our equations to those of Sec. 2 in \cite{EvE}. The latter
must be adapted to the particular imperfect fluid defined by Eq.
(\ref{Tij}).} In particular, we employ a combination of
propagation and constraint equations. These are:

{\bf (i)} The fluid conservation laws, represented by the
continuity equation
\begin{equation}
\dot{\mu}=-\mu(1+w)\Theta\,  \label{edc}
\end{equation}
and by Euler's formula
\begin{equation}
\mu\left(1+w+{\textstyle{2\over3}}h\right)\dot{u}_a=-Y_a
-\varepsilon_{abc}H^b{\rm curl}H^c-\dot{u}^b\Pi_{ba}\,,
\label{mdc}
\end{equation}
where $w\equiv p/\mu$, $h\equiv H^2/\mu$ and ${\rm
curl}H_a\equiv\varepsilon_{abc}\tilde{\nabla}^bH^c$. Generally,
$w$ is allowed to vary with
\begin{equation}
\dot{w}=-(1+w)\left(c_{\rm s}^2-w\right)\Theta\,,  \label{dotw}
\end{equation}
where $c_{\rm s}^2\equiv\dot{p}/\dot{\mu}$. Therefore, when
$\dot{w}=0$, then $c_{\rm s}^2=w=constant$. Note the total absence
of magnetic terms in Eq. (\ref{edc}), since the field's energy
density is conserved separately as a consequence of Maxwell's
equations.

{\bf (ii)} The kinematic propagation equations, comprising
Raychaudhuri's formula
\begin{equation}
\dot{\Theta}=-{\textstyle{1\over3}}\Theta^2-
{\textstyle{1\over2}}\mu\left(1+3w+h\right)-2\sigma^2
+A+\Lambda\,,  \label{Ray}
\end{equation}
where $A=\tilde{\nabla}^a\dot{u}_a$ (to first order) and $\Lambda$
is the cosmological constant, the vorticity propagation equation
\begin{equation}
\dot{\omega}_a=-{\textstyle{2\over3}}\Theta\omega_a+
\sigma_{ab}\omega^b-{\textstyle{1\over2}}{\rm curl}\dot{u}_a
\label{doto}
\end{equation}
and the shear propagation equation
\begin{equation}
\dot{\sigma}_{\langle ab\rangle}=
-{\textstyle{2\over3}}\Theta\sigma_{ab}- \sigma_{c\langle
a}\sigma^c{}_{b\rangle}+ \tilde{\nabla}_{\langle
a}\dot{u}_{b\rangle}+ {\textstyle{1\over2}}\Pi_{ab}-E_{ab}\,.
\label{dots}
\end{equation}
The above are supplemented by an equal number of constraints. In
the nomenclature of \cite{EvE}, these are the ($0\alpha$)-equation
\begin{equation}
{\textstyle{2\over3}}\tilde{\nabla}_a\Theta=\tilde{\nabla}^b\sigma_{ab}+
{\rm curl}\omega_a\,, \label{C2}
\end{equation}
the vorticity-divergence constraint
\begin{equation}
\tilde{\nabla}^a\omega_a=0\ \label{C3}
\end{equation}
and the $H_{ab}$-equation
\begin{equation}
H_{ab}=({\rm curl}\sigma)_{ab}+\tilde{\nabla}_{\langle
a}\omega_{b\rangle}\,,  \label{C1}
\end{equation}
where $({\rm curl}S)_{ab}\equiv\varepsilon_{cd\langle
a}\tilde{\nabla}^cS^d{}_{b\rangle}$ for every PSTF tensor $S_{ab}$
\cite{M1}. Note that constraint (\ref{C3}) ensures that the
$\omega_a$ is a solenoidal vector to first order.

{\bf (iii)} The magnetic equations, obtained from the covariant
Maxwell equations \cite{E2}. For a perfectly conducting medium,
the latter decompose into one propagation equation
\begin{equation}
\dot{H}_{\langle a\rangle}=-{\textstyle{2\over3}}\Theta H_a+
\sigma_{ab}H^b+\varepsilon_{abc}H^b\omega^c\,,  \label{M1}
\end{equation}
namely the magnetic induction equation, and three constraints
\begin{equation}
\varepsilon_{abc}\dot{u}^bH^c+{\rm curl}H_a=J_{\langle
a\rangle}\,, \label{M2}
\end{equation}
\begin{equation}
2\omega_aH^a=\epsilon\,,  \label{M3}
\end{equation}
\begin{equation}
\tilde{\nabla}^aH_a=0\,, \label{M4}
\end{equation}
where $J_{\langle a\rangle}$ is the projected current density and
$\epsilon=-J_au^a$ is the charge density. According to Eq.
(\ref{M3}), an overall electrical neutrality (i.e. $\epsilon=0$)
ensures that the vorticity vector and the magnetic field are
orthogonal. Also, Eq. (\ref{M4}) guarantees that $H_a$ is a
solenoidal. Contracted with $H_a$, the induction equation provides
the conservation law of the magnetic energy density
\begin{equation}
\left(H^2\right)^{.}=-{\textstyle{4\over3}}\Theta H^2-
2\sigma^{ab}\Pi_{ab}\,,  \label{medc}
\end{equation}
where the background anisotropy has added the last term of the
right hand side. Thus, unlike perturbed FRW models, the magnetic
energy density no longer drops as $S^{-4}$. Also, from Eq.
(\ref{cR(ij)}) bellow we obtain the expression
\begin{equation}
\Pi_{ab}={\textstyle{2\over3}}\Theta\sigma_{ab}- 2\sigma_{c\langle
a}\sigma^c{}_{b\rangle}-2E_{ab}+ {\cal R}_{\langle ab\rangle}\,,
\label{lPi}
\end{equation}
that connects $\Pi_{ab}$ to the rest of the linear anisotropic
sources.

Note that, so far, all equations have been derived from the fully
non-linear relations given in \cite{TB1}. Next we provide a new
pair of linear equations that govern the behaviour of $\Pi_{ab}$.
In particular, the time derivative of (\ref{Pi}) and Eqs.
(\ref{M1}) and (\ref{medc}) lead to the propagation equation of
the anisotropic magnetic stresses
\begin{equation}
\dot{\Pi}_{\langle ab\rangle}=
-{\textstyle{4\over3}}\Theta\Pi_{ab}+2\Pi_{c\langle
a}\left(\sigma^c{}_{b\rangle}+\varepsilon^{cd}{}{}_{b\rangle}\omega_d\right)-
{\textstyle{2\over3}}\mu h\sigma_{ab}\,. \label{dotPi}
\end{equation}
Similarly, by projecting the gradient of (\ref{Pi}) orthogonal to
the fluid flow we arrive at the constraint
\begin{equation}
\tilde{\nabla}^b\Pi_{ab}=\varepsilon_{abc}H^b{\rm curl}H^c-
{\textstyle{1\over6}}\tilde{\nabla}_aH^2\,.  \label{Picon}
\end{equation}

{\bf (iv)} The equations that determine the geometry of the
observer's rest space. The latter is characterised by the
3-Riemann tensor ${\cal R}_{abcd}$, which when contracted provides
the spatial Ricci tensor ${\cal R}_{ab}={\cal
R}^c_{\hspace{1mm}acb}$ and Ricci scalar ${\cal R}={\cal
R}^a_{\hspace{1mm}a}$.\footnote{We remind the reader that in
rotating spacetimes there are no proper spacelike hypersurfaces
orthogonal to the fluid flow. Here, we use the term `spatial',
when referring to the instantaneous rest space of a fundamental
observer, for convenience only. Also, when $\omega_a\neq0$, ${\cal
R}_{abcd}$ does not share the full symmetries of the spacetime
Riemann tensor \cite{EBH}.} For our purposes, the key expression
is the Gauss-Godazzi equation
\begin{equation}
{\cal R}_{ab}={\textstyle{1\over3}}{\cal R}h_{ab}+
{\textstyle{1\over2}}\Pi_{ab}-
{\textstyle{1\over3}}\Theta\left(\sigma_{ab}+\omega_{ab}\right)+
\sigma_{c\langle a}\sigma^c{}_{b\rangle}+
2\sigma_{c[a}\omega^c{}_{b]}+E_{ab}\,, \label{cRij}
\end{equation}
implying that the 3-Ricci tensor has also an antisymmetric part.
In fact, ${\cal R}_{ab}$ decomposes into
\begin{equation}
{\cal R}_{(ab)}={\textstyle{1\over3}}{\cal R}h_{ab}+
{\textstyle{1\over2}}\Pi_{ab}-
{\textstyle{1\over3}}\Theta\sigma_{ab}+ \sigma_{c\langle
a}\sigma^c{}_{b\rangle}+E_{ab}\,  \label{cR(ij)}
\end{equation}
and
\begin{equation}
{\cal R}_a=-{\textstyle{1\over3}}\Theta\omega_a-
\sigma_{ab}\omega^b\,, \label{cRi}
\end{equation}
where ${\cal R}_a\equiv{\textstyle{1\over2}}\varepsilon_{abc}{\cal
R}^{bc}$ represents the skew part of  ${\cal R}_{ab}$. The average
spatial curvature is determined by the sign of the 3-Ricci scalar.
The latter follows from the generalised Friedmann equation
\begin{equation}
{\cal R}=2\mu\left(1+{\textstyle{1\over2}}h\right)-
{\textstyle{2\over3}}\Theta^2+2\sigma^2+2\Lambda\,,  \label{cR}
\end{equation}
and evolves according to
\begin{equation}
\dot{{\cal R}}=-{\textstyle{2\over3}}\Theta{\cal R}+
2\left[\left(\sigma^2\right)^{.}+2\Theta\sigma^2\right]-
2\sigma^{ab}\Pi_{ab}-{\textstyle{4\over3}}\Theta A\,,
\label{dotcR}
\end{equation}

For the Weyl tensor components, we use Eqs. (\ref{dotPi}) and
(\ref{Picon}) to obtain a set of two propagation formulae, namely
the $\dot{E}$-equation
\begin{eqnarray}
\dot{E}_{\langle ab\rangle}&=&
-\Theta\left(E_{ab}-{\textstyle{1\over2}}\Pi_{ab}\right)+
3\left(\sigma_{c\langle
a}+{\textstyle{1\over3}}\varepsilon_{cd\langle a}\omega^d\right)
\left(E^c{}_{b\rangle}
-{\textstyle{1\over2}}\Pi^c{}_{b\rangle}\right)-
{\textstyle{1\over2}}\mu(1+w)\sigma_{ab} \nonumber\\&{}&+({\rm
curl}H)_{ab}\,, \label{ldotEij}
\end{eqnarray}
and the $\dot{H}$-equation
\begin{equation}
\dot{H}_{ab}=-\Theta H_{ab}+ 3\sigma_{c\langle
a}H^c{}_{b\rangle}-({\rm curl}E)_{ab}+{\textstyle{1\over2}}({\rm
curl}\,\Pi)_{ab}+ 2\varepsilon_{cd\langle
a}E^c{}_{b\rangle}\dot{u}^d\,, \label{dotHij}
\end{equation}
which are accompanied by an equal number of constraints, the
(${\rm div}E$)-equation
\begin{equation}
\tilde{\nabla}^bE_{ab}= \frac{\mu}{3S}\left({\cal
D}_a+{\textstyle{3\over4}}h{\cal B}_a\right)-
{\textstyle{1\over2}}\varepsilon_{abc}H^b{\rm curl}H^c+
\varepsilon_{abc}\sigma^b_{\hspace{1mm}d}H^{cd}\,, \label{Wecon}
\end{equation}
and the (${\rm div}H$)-equation
\begin{equation}
\tilde{\nabla}^bH_{ab}=
\mu\left(1+w+{\textstyle{2\over3}}h\right)\omega_a-
\varepsilon_{abc}\sigma^b_{\hspace{1mm}d}
\left(E^{cd}+{\textstyle{1\over2}}\Pi^{cd}\right)+
3\left(E_{ab}-{\textstyle{1\over6}}\Pi_{ab}\right)\omega^b\,,
\label{Wmcon}
\end{equation}
where ${\cal B}_a\equiv S\tilde{\nabla}_aH^2/H^2$. The latter is
an additional gauge-invariant variable representing
inhomogeneities in the magnetic energy density.

{\bf (v)} The linear propagation equations of the key
inhomogeneity variables ${\cal D}_a$, ${\cal Z}_a$ and ${\cal
M}_{ab}$. The former describes gradients in the fluid energy
density and evolves according to
\begin{eqnarray}
\dot{{\cal D}}_a&=&w\Theta{\cal D}_a-(1+w){\cal Z}_a+ \frac{\Theta
S}{\mu}\varepsilon_{abc}H^b{\rm curl}H^c+
{\textstyle{2\over3}}h\Theta S\dot{u}_a\nonumber\\&{}&-
\sigma_{ab}{\cal D}^b+\frac{\Theta S}{\mu}\Pi_{ab}\dot{u}^b\,,
\label{dotcDi}
\end{eqnarray}
obtained from Eq. (71) in \cite{TB1}. Note the last two terms in
the right hand side of the above, which reflect the background
anisotropy. Following Eq. (72) in \cite{TB1}, the expansion
gradients propagate as
\begin{eqnarray}
\dot{{\cal Z}}_a&=&-{\textstyle{2\over3}}\Theta{\cal Z}_a-
{\textstyle{1\over2}}\mu{\cal D}_a+SA_a-3S\sigma^2\dot{u}_a+
{\textstyle{3\over2}}S\varepsilon_{abc}H^b{\rm curl}H^c-
{\textstyle{1\over2}}\mu h{\cal B}_a\nonumber\\&{}&-
\sigma_{ab}{\cal Z}^b+ {\textstyle{3\over2}}S\Pi_{ab}\dot{u}^b-
2S\tilde{\nabla}_a\sigma^2\,, \label{dotcZi}
\end{eqnarray}
where, $A_a\equiv\tilde{\nabla}_aA$ and the last three terms are
due to the anisotropic background. Finally, inhomogeneities in the
magnetic field vector are governed by
\begin{eqnarray}
\dot{{\cal M}}_{ab}&=&-{\textstyle{2\over3}}\Theta{\cal M}_{ab}-
{\textstyle{2\over3}}H_a{\cal Z}_b- {\textstyle{1\over3}}\Theta
S\left(H_{[a}\dot{u}_{b]}+3H_{\langle a}\dot{u}_{b\rangle}\right)+
SH^c\tilde{\nabla}_b(\sigma_{ac}+\varepsilon_{acd}\omega^d)
\nonumber\\&{}&-S\varepsilon_{acd}H^cH^d{}_b+ \sigma_{ac}{\cal
M}^c{}_b- \sigma_{bc}{\cal M}_a{}^c+
SH^c\dot{u}_c\sigma_{ab}+2SH^c\sigma_{c[a}\dot{u}_{b]}\,,
\label{dotcMij}
\end{eqnarray}
which derives from Eq. (75) in \cite{TB1}. Here, the background
anisotropy is represented by the last four terms of the right hand
side.

%%%%%%%%%%%%%%%%%%%%%%%%%%%%%%%%%%
\section{The Bianchi~I Background}
%%%%%%%%%%%%%%%%%%%%%%%%%%%%%%%%%%
I has long been known that anisotropic exact Bianchi cosmologies
of type I can accommodate spatially homogeneous ordered magnetic
fields. We refer the reader to \cite{D}-\cite{C} for a sample of
such studies during the 60's and the early 70's. All these authors
assume, either directly or indirectly, that the magnetic field is
aligned along the shear eigenvector, and the same is also true for
our analysis. Note, however, that the field equations do not
necessarily impose any such condition upon the magnetic vector in
Bianchi~I models \cite{HJ}. In fact a magnetised Bianchi~I
universe, where the field is not a shear eigenvector, has been
studied by means of dynamical system techniques \cite{Lb}.

Here, we focus upon the anisotropic magnetic stresses and their
impact on cosmological perturbations. For our purposes, the whole
of the background anisotropy must result from the presence of the
field. This, in turn, will isolate the effects of the magnetic
stresses and facilitate their analysis.

%%%%%%%%%%%%%%%%%%%%%%%%%%%
\subsection{The Anisotropy}
%%%%%%%%%%%%%%%%%%%%%%%%%%%
The key effect of any large-scale magnetic field is the
introduction of a preferred direction, which is clearly manifested
by the eigenvalues of the $\Pi_{ab}$ tensor. According to Eq.
(\ref{Pi}), these eigenvalues are $-{\textstyle{2\over3}}H^2$
along the magnetic direction and ${\textstyle{1\over3}}H^2$ in the
orthogonal plane. In other words there is a negative pressure, a
{\em tension}, parallel to the field lines. Thus, we see the
prominent role played by the magnetic lines of force. Each small
flux tube is like a rubber band under tension, and infinitely
elastic \cite{Pa}. We will return to this property in sections 5
and 7, when discussing the field effects on the kinematics of the
universe and on the growth of density perturbations.

Given that $(\Pi_{ab}\Pi^{ab})^{1/2}\sim H^2$, the dimensionless
energy density ratio $h=H^2/\mu$ offers a particularly useful
measure of the magnetically induced anisotropy, provided that
matter is distributed isotropically. Clearly, the degree of
distortion inflicted on the background isotropy by the magnetic
presence depends on the field's strength relative to the fluid,
that is on the value of $h$ with respect to unity. When the field
is relatively weak, that is for $h\ll1$, one might treat the
background universe as isotropic to leading order. Strictly
speaking however, only anisotropic spacetimes can naturally
accommodate large-scale cosmological magnetic fields.

Let us assume that the entire background anisotropy is due to the
magnetic presence. This scheme may not represent the most general
case, but it helps to isolate the effects of the magnetic
stresses. Therefore, to zero order, both shear and Weyl
anisotropies result from the presence of the field.
Mathematically, we achieve this by demanding that the background
magnetic field is aligned in the direction of a shear eigenvector.
\begin{equation}
\sigma_{ab}H^b={\textstyle{2\over3}}\lambda H_a\,,  \label{sev}
\end{equation}
where $\tilde{\nabla}_a\lambda=0$ to zero order. Now, we introduce
a shear eigenframe with the background field vector along its
$x^1$-axis. In this frame, hereafter referred to as the magnetic
frame, both $\sigma_{ab}$ and $\Pi_{ab}$ are simultaneously
diagonalisable (recall that $H_a$ is the natural eigenvector of
$\Pi_{ab}$). Axial symmetry then implies
\begin{equation}
\sigma_{ab}=-\frac{\lambda}{H^2}\Pi_{ab}\,,  \label{sPi}
\end{equation}
thus guaranteeing that shear anisotropies result directly from the
magnetic stresses. The above resembles the Navier-Stokes equation,
so that $H^2/2\lambda$ may be thought of as the magneto-shear
viscocity coefficient.\footnote{The direct dependence of the
shear, and subsequently of the electric Weyl tensor, on the
magnetic stresses holds in the background only. No relation
analogous to (\ref{sev}) or (\ref{sPi}), between any two
ansisotropic sources, is ever imposed in the perturbed spacetime.}
Following Eq. (\ref{sev}), the shear eigenvalues are
${\textstyle{2\over3}}\lambda$, parallel to the field vector, and
$-{\textstyle{1\over3}}\lambda$ orthogonal to it. Also, relations
(\ref{sev}) and (\ref{sPi}) imply that
\begin{equation}
\sigma^2\equiv{\textstyle{1\over2}}\sigma_{ab}\sigma^{ab}
={\textstyle{1\over3}}\lambda^2\,, \label{lambda}
\end{equation}
to zero order.

We monitor the impact of the shear, relative to the average volume
expansion, via the dimensionless  shear-impact parameter
$\zeta\equiv\lambda/\Theta$. The latter measures the degree of the
kinematic anisotropy. As an example, consider the expansion rate
along the field lines. Then, on using (\ref{sev}) we find
\begin{equation}
\Theta_{ab}n^an^b={\textstyle{1\over3}}(1+2\zeta)\Theta\,,
\label{Thi}
\end{equation}
where $\Theta_{ab}\equiv\sigma_{ab}+{\textstyle{1\over3}}\Theta
h_{ab}$ is the zero-order expansion tensor and $n_a\equiv
H_a/\sqrt{H^2}$ is the unit vector in the direction of the field.
Although the shear effect depends on the specific form of $\zeta$,
it becomes insignificant when $\zeta\ll1$. The same condition
guarantees that the shear impact is also weak normally to the
field lines. In fact, result (\ref{lambda}) ensures that the shear
effect, relative to the volume expansion, is always negligible
provided $\zeta\ll1$.\footnote{The overall weakness of the shear
is also necessary, if the comoving gradients that describe the
inhomogeneities are to maintain the desirable geometrical and
physical interpretation they have within an almost-FRW universe
(see Appendix).}

%%%%%%%%%%%%%%%%%%%%%%%%%%
\subsection{The Evolution}
%%%%%%%%%%%%%%%%%%%%%%%%%%
In this section we discuss key features of the background
Bianchi~I spacetime, assuming that the magnetic field is the sole
source of its anisotropy. Unless stated otherwise, the equations
presented here are obtained from the linearised formulae of Sec. 3
by keeping their zero-order terms only.

\subsubsection{Energy Conservation}
%%%%%%%%%%%%%%%%%%%%%%%%%%%%%%%%%%%
The fluid and the magnetic energy densities obey the familiar
conservation laws. These are respectively represented by the
equations of continuity
\begin{equation}
\dot{\mu}=-(1+w)\mu\Theta\,,  \label{bedc}
\end{equation}
for the fluid, and
\begin{equation}
\left(H^2\right)^{.}=-{\textstyle{4\over3}}(1-\zeta)\Theta H^2\,,
\label{bH2}
\end{equation}
for the field. The latter derives from Eq. (\ref{medc}) on using
condition (\ref{sev}).

\subsubsection{Volume Expansion}
%%%%%%%%%%%%%%%%%%%%%%%%%%%%%%%%
To zero-order, result (\ref{lambda}) simplifies Raychaudhuri's
equation (\ref{Ray}) to
\begin{equation}
\dot{\Theta}=
-{\textstyle{1\over3}}\left(1+2\zeta^2\right)\Theta^2-
{\textstyle{1\over2}}\mu(1+3w+h)+ \Lambda\,,  \label{bRay}
\end{equation}
where the magnetic and the shear effects, respectively represented
by $h$ and $\zeta$, assist the matter in focusing the fluid
flow-lines. Similarly to the FRW models, we may introduce an
average scale factor ($S$) via
$\dot{S}/S={\textstyle{1\over3}}\Theta$, which also defines the
mean Hubble rate.

\subsubsection{Shear Anisotropies}
%%%%%%%%%%%%%%%%%%%%%%%%%%%%%%%%%%
The spatial flatness of the Bianchi~I spacetime and the
Gauss-Godazzi equation (see Eq. (83) in \cite{TB1}) provide an
extra evolution equation for the background shear
\begin{equation}
\dot{\sigma}_{ab}=-\Theta\sigma_{ab}+\Pi_{ab}\,,  \label{bdotsx}
\end{equation}
thus supplementing the standard zero-order propagation formula for
the shear. The latter is obtained from Eq. (\ref{dots}) by means
of condition (\ref{sPi}) and has the form
\begin{equation}
\dot{\sigma}_{ab}=-{\textstyle{2\over3}}
\left(1+{\textstyle{1\over2}}\zeta\right)\Theta\sigma_{ab}+
{\textstyle{1\over2}}\Pi_{ab}-E_{ab}\,,  \label{bdots}
\end{equation}
which guarantees that the zero-order shear evolution is shaped by
magnetic anisotropies and by long-range gravitational forces. The
flatness of the unperturbed spatial sections and Eq. (\ref{cR})
also supply a zero-order expression for the magnitude of the shear
tensor
\begin{equation}
\sigma^2=-\mu\left(1+{\textstyle{1\over2}}h\right)+
{\textstyle{1\over3}}\Theta^2-\Lambda\,,  \label{Fe}
\end{equation}
which is simply the background Friedmann equation. Note that on
using Eq. (\ref{lambda}) the above writes as
\begin{equation}
\mu\left(1+{\textstyle{1\over2}}h\right)
-{\textstyle{1\over3}}\Theta^2\left(1-\zeta^2\right) +\Lambda=0\,,
\label{Fe2}
\end{equation}
an expression that will prove useful later. Finally, employing
(\ref{sPi}) and (\ref{lambda}), the zero-order component of Eq.
(\ref{dotcR}) gives
\begin{equation}
\dot{\sigma}=-\Theta\sigma-{\textstyle{1\over\sqrt{3}}}\,\mu h\,,
\label{bdotsigma}
\end{equation}
which, on using (\ref{lambda}), is recast as
\begin{equation}
\dot{\lambda}=-\Theta\lambda-\mu h\,,  \label{dotl}
\end{equation}
that monitors the evolution of the $\lambda$-parameter.

\subsubsection{Magnetic Stresses}
%%%%%%%%%%%%%%%%%%%%%%%%%%%%%%%%%
Equations (\ref{bdotsx}) and (\ref{bdots}) combine to provide the
zero-order expression
\begin{equation}
\Pi_{ab}= {\textstyle{2\over3}}(1-\zeta)\Theta\sigma_{ab}-
2E_{ab}\,,  \label{bPi}
\end{equation}
which is also obtained by substituting (\ref{sPi}) into the
background component of Eq. (\ref{lPi}). The above provides a
unique relation between $\Pi_{ab}$, $\sigma_{ab}$ and $E_{ab}$.
This fine-tuning of the unperturbed anisotropic sources results
from the spatial flatness of the Bianchi~I model and ensures that
Eqs. (\ref{bdotsx}), (\ref{bdots}) are simultaneously satisfied.
Using condition (\ref{sPi}), Eqs. (\ref{dotPi}) gives
\begin{equation}
\dot{\Pi}_{ab}=-{\textstyle{4\over3}}(1-\zeta)\Theta\Pi_{ab}\,,
\label{bdotPi}
\end{equation}
to zero order. The latter is also obtained from Eq. (\ref{sPi}) by
means of (\ref{sPi}), (\ref{bH2}) and (\ref{bdotsigma}).

\subsubsection{Weyl anisotropies}
%%%%%%%%%%%%%%%%%%%%%%%%%%%%%%%%%
In the unperturbed universe, Eq. (\ref{bPi}) guarantees the direct
dependence of the Weyl anisotropies on the magnetic stresses. In
fact, Eqs. (\ref{sev}) and (\ref{bPi}) confirm that $H_a$ is also
an eigenvector of $E_{ab}$, while Eqs. (\ref{sPi}) and (\ref{bPi})
ensure that
\begin{equation}
E_{ab}=-\frac{\eta}{H^2}\Pi_{ab}\,,  \label{eWPi}
\end{equation}
where $\eta\equiv{\textstyle{1\over2}}\mu
h+{\textstyle{1\over3}}(1-\zeta)\lambda\Theta$ has the dimensions
of $E_{ab}$. The background evolution of the electric Weyl tensor
is given by the zero-order part of Eq. (\ref{ldotEij})
\begin{equation}
\dot{E}_{\langle ab\rangle}=
-\Theta\left(E_{ab}-{\textstyle{1\over2}}\Pi_{ab}\right)+
3\sigma_{c\langle a} \left(E^c{}_{b\rangle}
-{\textstyle{1\over2}}\Pi^c{}_{b\rangle}\right)-
{\textstyle{1\over2}}\mu(1+w)\sigma_{ab}\,. \label{bdotEij}
\end{equation}
Hence, condition (\ref{sev}) and axial symmetry guarantee the
direct dependence of the whole background anisotropy on the
magnetic presence. Once the parameter $\zeta$ has been
established, the zero-order evolution of both $\sigma_{ab}$ and
$E_{ab}$ is determined by Eq. (\ref{bdotPi}).

%%%%%%%%%%%%%%%%%%%%%%%%%%%%%%%
\subsection{The Alfv\'en Speed}
%%%%%%%%%%%%%%%%%%%%%%%%%%%%%%%
Disturbances in a magnetised medium propagate via the Alfv\'en
speed. For an arbitrarily strong magnetic field we define
\begin{equation}
c_{\rm a}^2\equiv\frac{h}{1+w+h}\,,  \label{Alf}
\end{equation}
so that $c_{\rm a}^2<1$ always. This coincides with the
definitions in \cite{TB1} and \cite{TB2} for a weak field in a
dust-dominated universe (i.e. $h\ll1$ and $w\equiv0$), but
deviates when the fluid has non-vanishing pressure. The reader
must take this difference into account when comparing our
pre-recombination equations to those in \cite{TB1} or \cite{TB2}.

On using Eqs. (\ref{bedc}) and (\ref{bH2}), we obtain the
evolution law for the ratio between the magnetic and the fluid
energy densities
\begin{equation}
\dot{h}=-{\textstyle{1\over3}}(1-3w-4\zeta)\Theta h\,.
\label{doth}
\end{equation}
Note that when $w\neq1/3$ and the field is weak (i.e.
$\zeta\ll1$), the last term in the right hand side of (\ref{doth})
is negligible. Hence, in the dust era, $h$ evolves exactly as in
weakly magnetised FRW cosmologies. In the radiation era, however,
$w=1/3$ and therefore $\dot{h}=4\lambda h/3$. In this case the
behaviour of $h$ depends on the specific form of $\lambda$, that
is on the relation between $\sigma_{ab}$ and $\Pi_{ab}$. This
complication, which was first pointed out in \cite{Z}, corresponds
to the `critical case' discussed in \cite{B1,BM}.

The Alfv\'en speed also shows the same subtle behaviour during the
radiation epoch. Indeed, combining definition (\ref{Alf}) with
Eqs. (\ref{dotw}) and (\ref{doth}) we obtain
\begin{equation}
\left(c_{\rm a}^2\right)^.=-{\textstyle{1\over3}}\left(1-3c_{\rm
s}^2-4\zeta\right)\left(1-c_{\rm a}^2\right)\Theta c_{\rm a}^2\,.
\label{dotAlf}
\end{equation}
Clearly, when $c_{\rm s}^2=1/3$, the evolution of the Alfv\'en
speed couples to the shear impact parameter, even for $c_{\rm
a}^2\ll1$.

%%%%%%%%%%%%%%%%%%%%%%%%%%%%%%%%%%%%%%%%
\section{The Case of a Barotropic Fluid}
%%%%%%%%%%%%%%%%%%%%%%%%%%%%%%%%%%%%%%%%
The set of equations (\ref{dotcDi})-(\ref{dotcMij}), which holds
for a general perfect fluid with pressure, does not contain an
evolution formula for the pressure gradients. The reason being
that the propagation of $Y_a$ is determined directly form
(\ref{dotcDi}), once the material content of the universe has been
specified. Here we consider a universe filled with a single
barotropic perfect fluid. This has the simple equation of state
$p=p(\mu)$ which implies that
\begin{equation}
SY_a=\mu c_{\rm s}^2{\cal D}_a\,.  \label{YiDi}
\end{equation}

%%%%%%%%%%%%%%%%%%%%%%%%%%%%%%%%%
\subsection{The Scalar Variables}
%%%%%%%%%%%%%%%%%%%%%%%%%%%%%%%%%
The gauge-invariant variables defined next, allow to distinguish
between inhomogeneities of different types. To begin with, we
decompose the comoving spatial gradient of ${\cal D}_a$ as
\begin{equation}
S\tilde{\nabla}_b{\cal
D}_a\equiv\Delta_{ab}={\textstyle{1\over3}}\Delta
h_{ab}+\varepsilon_{abc}W^c+\Sigma_{ab}\,,  \label{Delij}
\end{equation}
where $\Delta\equiv\Delta^a_{\hspace{1mm}a}$,
$W_a\equiv{\textstyle{1\over2}}\varepsilon_{abc}\Delta^{bc}$ and
$\Sigma_{ab}\equiv\Delta_{\langle ab\rangle}$. The scalar $\Delta$
describes the average gravitational clumping of the fluid, the
vector $W_a$ represents rotational instabilities and $\Sigma_{ab}$
describes volume-true infinitesimal distortions in clustering
\cite{EBH,MTM}. We also apply similar decompositions to ${\cal
Z}_{ab}\equiv S\tilde{\nabla}_b{\cal Z}_a$, for the expansion
inhomogeneities, and to ${\cal B}_{ab}\equiv
S\tilde{\nabla}_b{\cal B}_a$, which is associated with gradients
in the magnetic energy density.

Given that density perturbations (i.e. scalar matter aggregations)
are the focus of this analysis the following scalars are crucial
\begin{equation}
\Delta\equiv\Delta^a_{\hspace{1mm}a}=
\frac{S^2}{\mu}\tilde{\nabla}^2\mu\,,\hspace{10mm}{\cal
Z}\equiv{\cal
Z}^a_{\hspace{1mm}a}=S^2\tilde{\nabla}^2\Theta\,\hspace{10mm}{\rm
and}\hspace{10mm}{\cal B}\equiv{\cal
B}^a_{\hspace{1mm}a}=\frac{S^2}{H^2}\tilde{\nabla}^2H^2\,,
\label{scalars}
\end{equation}
where the equalities hold in the linear regime only. The above
variables, which are all gauge-invariant, are supplemented by the
first-order scalar
\begin{equation}
{\cal K}\equiv S^2{\cal R}\,,  \label{cK}
\end{equation}
representing spatial curvature perturbations. Note that only
$\Delta$, ${\cal B}$ and ${\cal K}$ are dimensionless, while
${\cal Z}$ has inverse time dimensions.

%%%%%%%%%%%%%%%%%%%%%%%%%%%%%%%%%%%%
\subsection{The Kinematic Evolution}
%%%%%%%%%%%%%%%%%%%%%%%%%%%%%%%%%%%%
\subsubsection{Acceleration}
%%%%%%%%%%%%%%%%%%%%%%%%%%%%
Employing the barotropic relation (\ref{YiDi}), the momentum
density conservation law (see Eq. (\ref{mdc})) gives
\begin{equation}
\dot{u}_a=-\frac{c_{\rm s}^2}{(1+w)S}
\left(1-{\textstyle{2\over3}}c_{\rm a}^2\right){\cal D}_a
-\frac{c_{\rm a}^2}{\mu h}\varepsilon_{abc}H^b{\rm curl}H^c
+\frac{c_{\rm s}^2c_{\rm a}^2}{\mu h(1+w)S}\Pi_{ab}{\cal D}^b\,,
\label{bai}
\end{equation}
with the background anisotropy reflected in the last term on the
right. Now, expression (\ref{Pi}) recasts Eq. (\ref{bai}) into
\begin{equation}
\dot{u}_a=-\frac{c_{\rm s}^2}{(1+w)S}\left(1-c_{\rm
a}^2\right){\cal D}_a- \frac{c_{\rm a}^2}{\mu
h}\varepsilon_{abc}H^b{\rm curl}H^c- \frac{c_{\rm s}^2c_{\rm
a}^2}{(1+w)S}n_an^b{\cal D}_b\,, \label{mbai}
\end{equation}
where the last term acts only along the direction of the field and
the second last is transverse. Hence, the acceleration components
are
\begin{equation}
\dot{u}_a=\left\{\begin{array}{l} -\frac{c_{\rm s}^2}{(1+w)S}{\cal
D}_a\hspace{10mm}{\rm parallel~to}\,H_a\,,\\\\-\frac{c_{\rm
s}^2}{(1+w)S}\left(1-c_{\rm a}^2\right){\cal D}_a- \frac{c_{\rm
a}^2}{\mu h}\varepsilon_{abc}H^b{\rm curl}H^c\hspace{10mm}{\rm
orthogonal~to} \,H_a\,,\\
\end{array}\right.  \label{ba123}
\end{equation}
confirming that the field does not affect the fluid motion along
its lines of force. Note that, when the field is weak (i.e. for
$c_{\rm a}^2\ll1$) the acceleration components reduce to their FRW
counterparts (compare to Eq. (31) in \cite{TB2}).

Equation (\ref{bai}) verifies that, in the absence of fluid
pressure (i.e. when $c_{\rm s}^2\equiv0$), $\dot{u}_a$ depends
only on magnetic gradients and is always normal to the field
vector. Clearly, this is not the case when $p\neq0$. For
non-vanishing pressure, one might also say that $\dot{H}_a$ is no
longer normal to the fluid flow. We verify these statements by
simply contracting (\ref{bai}) with $H_a$. We find that
\begin{equation}
H^a\dot{u}_a=\dot{H}^au_a=-\frac{c_{\rm s}^2}{(1+w)S}H^a{\cal
D}_a\,, \label{bHiai}
\end{equation}
where generally $H^a{\cal D}_a\neq0$. The above condition, which
is independent of the field's strength, is identical to that
obtained by the almost-FRW treatment \cite{TB2}.

According to the kinematic equations (\ref{Ray})-(\ref{dots}), the
acceleration gradients affect all aspects of the expansion. They
also have an impact on the spatial curvature (see Eq.
(\ref{dotcR})). Consequently, the following linear decomposition
is particularly useful
\begin{eqnarray}
\tilde{\nabla}_b\dot{u}_a&=& -\frac{c_{\rm s}^2}{(1+w)S^2}
\left(1-{\textstyle{2\over3}}c_{\rm a}^2\right)\Delta_{ab}-
\frac{c_{\rm a}^2}{2S^2}{\cal B}_{ab}+ \frac{c_{\rm a}^2}{\mu h
S}H^c\tilde{\nabla}_c{\cal M}_{ab}-
{\textstyle{4\over9}}(1-\zeta)c_{\rm a}^2\Theta
\varepsilon_{abc}\omega^c\nonumber\\&{}&+
{\textstyle{1\over3}}c_{\rm a}^2{\cal R}_{ab}+ \frac{c_{\rm
s}^2c_{\rm a}^2}{\mu h(1+w)S^2} \Pi_a{}^{c}\Delta_{cb}-
\frac{4c_{\rm a}^2\Theta}{3\mu h}(1-\zeta)
\Pi_a{}^{c}\varepsilon_{bcd}\omega^d- \frac{c_{\rm a}^2}{\mu
h}\Pi^{cd}{\cal R}_{acbd}\,,  \label{bGjai}
\end{eqnarray}
where the last three terms result from the anisotropic background.
Note the magneto-curvature contributions in the right hand side of
the above. Their presence manifests the vectorial nature of the
magnetic field and underlines the coupling between magnetism and
geometry. Using Eq. (\ref{Pi}), the two magneto-curvature terms
reduce to one given by $c_{\rm a}^2H^cH^d{\cal R}_{acbd}/\mu h$.
The latter represents a force transverse to the background field
lines, as the symmetries of ${\cal R}_{abcd}$ confirm. This term,
which appears inevitably when the 3-gradients of the field vector
commute (see Eq. (B9) in \cite{Du}), can trigger a range of rather
unexpected effects.

Next the trace, the skew part and the symmetric trace-free
component of Eq. (\ref{bGjai}) will be used to analyse the
magnetohydrodynamical effects upon the kinematics, the dynamics
and the geometry of our cosmological model.

\subsubsection{Deceleration Parameter}
%%%%%%%%%%%%%%%%%%%%%%%%%%%%%%%%%%%%%%
On using expression (\ref{Pi}), the linearised trace of
(\ref{bGjai}) gives
\begin{equation}
A=-\frac{c_{\rm s}^2}{(1+w)S^2}\left[\Delta_{ab}n^an^b
+\left(1-c_{\rm a}^2\right)\Delta_{ab}v^{ab}\right]- \frac{c_{\rm
a}^2}{2S^2}{\cal B}+ c_{\rm a}^2{\cal R}_{ab}n^an^b\,, \label{bA}
\end{equation}
where $A=\tilde{\nabla}^a\dot{u}_a$ to first order (see Eq.
(\ref{Ray})) and $v_{ab}\equiv h_{ab}-n_an_b$ projects into the
2-plane orthogonal to $H_a$. The scalars $\Delta_{ab}n^an^b$ and
${\cal R}_{ab}n^an^b$ respectively represent density and
3-curvature perturbations parallel to the background field lines.
On the other hand, $\Delta_{ab}v^{ab}$ describes matter
aggregations orthogonal to $H_a$, so that
$\Delta_{ab}n^an^b+\Delta_{ab}v^{ab}=\Delta$. Note that
$S_{ab}n^an^b+S_{ab}v^{ab}=S^a{}_a$ for any spacelike tensor
$S_{ab}$, while $S_{ab}n^an^b=S_{11}$ and
$S_{ab}v^{ab}=S_{22}+S_{33}$ in the magnetic frame.

Substituting the above result into Eq. (\ref{Ray}) we obtain the
Raychaudhuri equation for a magnetised almost Bianchi~I universe
filled with a perfectly conducting ideal fluid. The resulting
formula is then transformed into the following alternative
expression
\begin{eqnarray}
{\textstyle{1\over3}}\Theta^2{\rm q}&=&
{\textstyle{1\over2}}\mu(1+3w+h)+ \frac{c_{\rm
s}^2}{(1+w)S^2}\left[\Delta_{ab}n^an^b +\left(1-c_{\rm
a}^2\right)\Delta_{ab}v^{ab}\right] +\frac{c_{\rm a}^2}{2S^2}{\cal
B}\nonumber\\&{}&-c_{\rm a}^2{\cal R}_{ab}n^an^b+2\sigma^2-
\Lambda\,, \label{q}
\end{eqnarray}
where ${\rm q}\equiv-\ddot{S}S/\dot{S}^2$ is the dimensionless
deceleration parameter. Clearly, the sign of the right hand side
determines the state of the expansion, with positive terms slowing
it down and negative accelerating it. As expected, the field slows
the universe down by adding to the total energy density, and the
shear has a similar decelerating impact. Comparing to Eq. (36) in
\cite{TB2}, we see that the effect from local increases in the
magnetic energy density (i.e. for ${\cal B}>0$) remains unchanged.
However, the anisotropic background has left its signature through
a number of direction dependent effects. In particular, for
relatively strong magnetic fields (i.e. when $c_{\rm a}^2\sim1$),
matter aggregations contribute differently along the field
direction than perpendicularly to $H_a$. In fact, as the magnetic
field gets stronger (i.e. as $c_{\rm a}^2\rightarrow1$), the
decelerating effect from $\Delta_{ab}v^{ab}$ tends to zero.

According to Eq. (\ref{q}), only curvature perturbations parallel
to the magnetic force-lines have an influence on the expansion
rate. This effect was originally identified in \cite{TB2},
although there, the isotropy of the background FRW model meant
that it was independent of direction. Here, as in \cite{TB2},
negative curvature perturbations add to the total deceleration,
whereas a positive ${\cal R}_{ab}n^an^b$ tends to accelerate the
universe. This sounds odd at first, given that positive curvature
is traditionally associated with gravitational collapse. However,
the tension carried along the direction of the field, means that
small magnetic flux tubes behave like elastic rubber bands
\cite{Pa}. When coupled to geometry, this property of the field
gives rice to a curvature stress that opposes the action of the
magnetic energy density gradients. Hence, in Eq. (\ref{q}), {\em
the magneto-geometrical term reacts to local changes in the
spatial curvature by modifying the expansion rate of the perturbed
region accordingly}. Intuitively, the coupling between magnetism
and geometry has injected the elastic properties of the field into
space itself. Note that the effect of the relativistic
magneto-curvature term $c_{\rm a}^2{\cal R}_{ab}n^an^b$ in Eq.
(\ref{q}), bares a striking resemblance to the classical
(non-relativistic) curvature stress exerted by a distorted field.
Following \cite{Pa}, the tension from field lines with a local
radius of curvature $R$, exerts a transverse force $\sim H^2/R$
per unit volume. Here, the distortion in the field pattern is
triggered by perturbations in the spatial geometry itself.

Comparing to Eq. (36) in \cite{TB2}, we note that the
magneto-curvature effect identified in Eq. (\ref{q}) reduces to
that predicted by the FRW-based analysis when ${\cal
R}_{ab}n^an^b={\textstyle{1\over3}}{\cal R}$. Thus, for a weak
field, the two equations become indistinguishable when the
curvature perturbation along $H_a$ takes the (directional) average
value. Generally, however, ${\cal R}_{ab}n^an^b$ has a rather
complicated evolution that also depends on magnetic, shear and
Weyl contributions. In fact, contracting Eq. (\ref{cRij}) with the
field vector we find
\begin{equation}
{\cal R}_{ab}n^an^b={\textstyle{1\over3}}{\cal R}-
{\textstyle{1\over3}}\mu h- {\textstyle{2\over3}}\sigma^2-
{\textstyle{1\over3}}(1-2\zeta)\Theta\sigma_{ab}n^an^b+E_{ab}n^an^b\,,
\label{cR1}
\end{equation}
where $\sigma_{ab}n^an^b$ and $E_{ab}n^an^b$ are respectively the
components of the shear and the electric Weyl tensors parallel to
the filed lines.

The anisotropy of the Bianchi~I background has introduced
directional effects to the expansion dynamics, along and normally
to the background magnetic field lines. As we shall show next,
similar direction-dependent complications affect almost every
aspect of the evolution. None of these effects could have been
identified by an FRW-based treatment, where the isotropy of the
unperturbed spacetime allowed only the average of these effects to
be measured.

\subsubsection{Vorticity}
%%%%%%%%%%%%%%%%%%%%%%%%%
Substituting the background relation (\ref{sPi}) to the vorticity
propagation equation (\ref{doto}) we have
\begin{equation}
\dot{\omega}_a=
-{\textstyle{2\over3}}\left(1+{\textstyle{1\over2}\zeta}\right)
\Theta\omega_a- {\textstyle{1\over2}}{\rm curl}\dot{u}_a\,,
\label{bdoto1}
\end{equation}
where $\zeta$ always represents the magnetically induced kinematic
anisotropies. Using Eqs. (\ref{Pi}), (\ref{cRi}), (\ref{sev}),
(\ref{sPi}), (\ref{bH2}) and (\ref{Delij}), the commutator (B8)
from \cite{Du}, the symmetries of ${\cal R}_{abcd}$ (see eq. (81)
in \cite{EBH}) and the linear relation $W_a=-(1+w)\Theta
S^2\omega_a$, we obtain an expression for the skew part of Eq.
(\ref{bGjai}). Then, given that ${\rm
curl}\dot{u}_a=\varepsilon_{abc}\tilde{\nabla}^b\dot{u}^c$, the
vorticity propagation equation is recast as
\begin{equation}
\dot{\omega}_a+{\textstyle{2\over3}}
\left[\left(1+{\textstyle{1\over2}}\zeta\right)
-{\textstyle{3\over2}} \left(1-{\textstyle{1\over2}}c_{\rm
a}^2\right)c_{\rm s}^2\right]\Theta\omega_a=-\frac{c_{\rm
a}^2}{2\mu h}H^b\tilde{\nabla}_b{\rm curl}H_a+ \frac{c_{\rm
s}^2c_{\rm a}^2}{2\mu
h(1+w)S^2}\varepsilon_{abc}\Pi^{bd}\Sigma^c{}_d\,, \label{bdoto2}
\end{equation}
where $\Sigma_{ab}=(S^2/\mu)\tilde{\nabla}_{\langle
a}\tilde{\nabla}_{b\rangle}\mu$ to first order. The above holds
for an electrically neutral medium with $H^a\omega_a=0$. As
predicted by the almost-FRW treatment (see Eq. (37) in
\cite{TB2}), magnetism becomes a source of vorticity when ${\rm
curl}H_a$ varies in the direction of the field vector. Note that
before recombination distortions in the density distribution can
also generate vorticity. According to Eq. (\ref{bdoto2}), a
relatively strong magnetic field with $\zeta$, $c_{\rm a}^2\sim1$
can modify the vortex dilution caused by the volume expansion.

\subsubsection{Shear}
%%%%%%%%%%%%%%%%%%%%%
The symmetric trace-free part of (\ref{bGjai}) transforms Eq.
(\ref{dots}) into
\begin{eqnarray}
\dot{\sigma}_{\langle ab\rangle}&=&-
{\textstyle{2\over3}}\Theta\sigma_{ab}- \frac{c_{\rm
s}^2}{(1+w)S^2} \left(1-{\textstyle{2\over3}}c_{\rm
a}^2\right)\Sigma_{ab}- \frac{c_{\rm a}^2}{2S^2}{\cal B}_{\langle
ab\rangle}+ {\textstyle{1\over3}}c_{\rm a}^2{\cal R}_{\langle
ab\rangle}+ {\textstyle{1\over2}}\Pi_{ab}\nonumber\\&{}& +
\frac{c_{\rm a}^2}{\mu hS}H^c\tilde{\nabla}_c {\cal M}_{\langle
ab\rangle}- E_{ab}-\sigma_{c\langle a}\sigma^c{}_{b\rangle}+
\frac{c_{\rm s}^2c_{\rm a}^2}{\mu h (1+w)S^2} \Pi_{c\langle
a}\Delta^c{}_{b\rangle}\nonumber\\&{}&- \frac{c_{\rm a}^2}{\mu
h}\Pi_{cd} {\cal R}_{\langle
a\hspace{1mm}b\rangle}^{\hspace{2mm}c\hspace{2mm}d}- \frac{4c_{\rm
a}^2\Theta}{3\mu h}(1-\zeta) \Pi^c{}_{\langle
a}\varepsilon_{b\rangle cd}\omega^d\,. \label{bardots}
\end{eqnarray}
Comparing with Eq. (38) in \cite{TB2}, we identify the effects of
the background anisotropy in the last four terms. The extra
magnetic input to the $\Sigma_{ab}$-term, which is negligible for
relatively weak fields, results from the magnetic contribution to
the total energy density of the universe. Any other changes are
due to differences in the Alfv\'en speed definitions.

An additional useful relation is the propagation equation of the
shear magnitude. We obtain it by contracting Eq. (83) in
\cite{TB1} with $\sigma_{ab}$. We then employ results
(\ref{bGjai}) and (\ref{bA}), the symmetries of ${\cal R}_{abcd}$
and the constraint ${\cal M}^a_{\hspace{1mm}a}=0$ to find
\begin{eqnarray}
\left(\sigma^2\right)^{.}&=&-2\Theta\sigma^2-\mu
h\sigma_{ab}n^an^b- {\textstyle{1\over3}}\zeta\Theta
\left[2\left(1+{\textstyle{1\over2}}c_{\rm a}^2\right){\cal
R}_{ab}n^an^b -{\cal R}_{ab}v^{ab}\right] \nonumber\\&{}&-
\frac{\zeta c_{\rm s}^2\Theta}{3(1+w)S^2}\left[2\Delta_{ab}n^an^b-
\left(1-c_{\rm a}^2\right)\Delta_{ab}v^{ab}\right]
\nonumber\\&{}&- \frac{\zeta c_{\rm a}^2\Theta}{6S^2} \left(2{\cal
B}_{ab}n^an^b-{\cal B}_{ab}v^{ab}\right)\,, \label{bardotsigma}
\end{eqnarray}
where $\sigma_{ab}n^an^b$ satisfies Eq. (\ref{cR1}). Also, ${\cal
B}_{ab}n^an^b$ and ${\cal B}_{ab}v^{ab}$ respectively represent
magnetic energy density gradients parallel and orthogonal to the
background field lines. Similarly, ${\cal R}_{ab}v^{ab}$ is the
3-curvature perturbation in the plane normal to $H_a$. Note that
the anisotropic effects, manifested by the direction dependent
terms of the right hand side, remain even when the field is
relatively weak. The above result will prove useful in Sec. 5,
when studying spatial curvature perturbations.

%%%%%%%%%%%%%%%%%%%%%%%%%%%%%%%%%%
\subsection{The Dynamic Evolution}
%%%%%%%%%%%%%%%%%%%%$$$$%%%%%%%%%%
\subsubsection{Perturbations in the Matter Density}
%%%%%%%%%%%%%%%%%%%%%%%%%%%%%%%%%%%%%%%%%%%%%%%%%%%
In the linear regime, magnetised density gradients propagate
according to Eq. (\ref{dotcDi}). For a barotropic perfectly
conducting  medium, the latter becomes
\begin{eqnarray}
\dot{{\cal D}}_a&=& \left(w-c_{\rm s}^2c_{\rm
a}^2+{\textstyle{1\over3}}\zeta\right)\Theta{\cal D}_a- (1+w){\cal
Z}_a+ \frac{\Theta S}{\mu}\left(1-c_{\rm
a}^2\right)\varepsilon_{abc}H^b{\rm curl}H^c
\nonumber\\&{}&-\left(c_{\rm s}^2c_{\rm a}^2-\zeta\right)\Theta
n_an^b{\cal D}^b\,, \label{bdotcDi}
\end{eqnarray}
on using (\ref{Pi}), the zero-order relation (\ref{sPi}), the
barotropic expression (\ref{bai}) and result (\ref{bHiai}). We can
now see that the field affects density perturbations along as well
as normally to its own force-lines. Parallel to $H_a$, the density
gradients are influenced by the magnetically induced kinematical
anisotropies, with their presence manifested by the shear impact
parameter $\zeta$.

The evolution of linear matter aggregations is determined by the
3-divergence of Eq. (\ref{bdotcDi}). Using the commutation law
between time derivatives and 3-gradients (see formula (B18) in
\cite{Du}), we obtain
\begin{eqnarray}
\dot{\Delta}&=&
\left(w-{\textstyle{4\over3}}\zeta\right)\Theta\Delta_{ab}n^an^b+
\left(w-c_{\rm s}^2c_{\rm a}^2+{\textstyle{2\over3}}\zeta\right)
\Theta\Delta_{ab}v^{ab}- (1+w){\cal Z} \nonumber\\&{}&+
{\textstyle{1\over2}}(1+w)c_{\rm a}^2\Theta{\cal B}- (1+w)c_{\rm
a}^2\Theta{\cal K}_{ab}n^an^b\,. \label{bdotDel}
\end{eqnarray}
Note that different quantities contribute differently parallel
than perpendicular to the field lines. Also, the magneto-shear
effects, which are imprinted in the impact parameter $\zeta$,
vanish when $\Delta_{ab}n^an^b$ takes the directional average
value ${\textstyle{1\over3}}\Delta$.

\subsubsection{Perturbations in the Expansion}
%%%%%%%%%%%%%%%%%%%%%%%%%%%%%%%%%%%%%%%%%%%%%%
Starting from Eq. (\ref{dotcZi}) we employ expression (\ref{Pi}),
the zero-order relations (\ref{sPi}), (\ref{sev}) and (\ref{Fe}),
together with the linear equations (\ref{bai}), (\ref{bHiai}) and
(\ref{bA}) to arrive at the propagation formula of the expansion
gradients in a barotropic fluid environment
\begin{eqnarray}
\dot{{\cal
Z}}_a&=&-{\textstyle{2\over3}}\left(1-{\textstyle{1\over2}}\zeta\right)\Theta{\cal
Z}_a -{\textstyle{1\over2}}\mu\left[1+4c_{\rm s}^2c_{\rm
a}^2\left(1+{\textstyle{3\over2}}r\right)\right]{\cal
D}_a\nonumber\\&{}&-{\textstyle{1\over2}}\mu h{\cal B}_a+
{\textstyle{3\over2}}\left[1-{\textstyle{4\over3}}c_{\rm
a}^2\left(1+{\textstyle{3\over2}}r\right)\right]S\varepsilon_{abc}H^b{\rm
curl}H^c\nonumber\\&{}& -\frac{c_{\rm
s}^2}{(1+w)S}\tilde{\nabla}_a\left[\Delta_{ab}n^an^b+\left(1-c_{\rm
a}^2\right)\Delta_{ab}v^{ab}\right] -\frac{c_{\rm
a}^2}{2S}\tilde{\nabla}_a{\cal B}\nonumber\\&{}& -\zeta\Theta
n_an^b{\cal Z}_b+\frac{3\mu hc_{\rm
s}^2}{2(1+w)}\left[1-{\textstyle{4\over3}}c_{\rm
a}^2\left(1+{\textstyle{3\over2}}r\right)\right]n_an^b{\cal
D}_b\nonumber\\&{}& +c_{\rm a}^2S\tilde{\nabla}_a\left({\cal
R}_{ab}n^an^b\right)-2\zeta\Theta
S\tilde{\nabla}_a\left(\sigma_{ab}n^an^b\right)\,, \label{bdotcZi}
\end{eqnarray}
where the dimensionless parameter
$r\equiv\frac{1}{h}\left[1-\frac{1}{\mu}\left({\textstyle{1\over3}}\Theta^2-
\Lambda\right)\right]$ will prove useful later.

Taking the comoving spatial divergence of Eq. (\ref{bdotcZi}),
using the commutator between time derivatives and spatial
gradients of first order spacelike vectors (see Eq. (B18) in
\cite{Du}), together with expression (\ref{Pi}) and the background
relation (\ref{sPi}) we find that, to first order,
\begin{eqnarray}
\dot{{\cal Z}}&=&
-{\textstyle{2\over3}}\Theta\left[(1+2\zeta){\cal Z}_{ab}n^an^b
+(1-\zeta){\cal Z}_{ab}v^{ab}\right]\nonumber\\&{}&
-{\textstyle{1\over2}}\mu\left\{\left[1+\frac{c_{\rm
s}^2h}{1+w}(1+6r)\right]\Delta_{ab}n^an^b+\left[1+4c_{\rm
s}^2c_{\rm a}^2\left(1+{\textstyle{3\over2}}r\right)\right]
\Delta_{ab}v^{ab}\right\}\nonumber\\&{}& +{\textstyle{1\over4}}\mu
h\left[1-4c_{\rm
a}^2\left(1+{\textstyle{3\over2}}r\right)\right]{\cal B}
-{\textstyle{3\over2}}\mu h\left[1-{\textstyle{4\over3}}c_{\rm
a}^2\left(1+{\textstyle{3\over2}}r\right)\right]{\cal
K}_{ab}n^an^b \nonumber\\&{}&- \frac{c_{\rm
s}^2}{1+w}\tilde{\nabla}^2\left[\Delta_{ab}n^an^b+\left(1-c_{\rm
a}^2\right)\Delta_{ab}v^{ab}\right] -{\textstyle{1\over2}}c_{\rm
a}^2\tilde{\nabla}^2{\cal B}\nonumber\\&{}& +c_{\rm
a}^2\tilde{\nabla}^2\left({\cal K}_{ab}n^an^b\right)- 2\zeta\Theta
S^2\tilde{\nabla}^2\left(\sigma_{ab}n^an^b\right)\,,
\label{bdotcZ}
\end{eqnarray}
where ${\cal Z}_{ab}n^an^b$ and ${\cal Z}_{ab}v^{ab}$ are
respectively the expansion gradients parallel and orthogonal to
the zero-order magnetic vector. Note that most of the anisotropic
effects in (\ref{bdotcZi}), (\ref{bdotcZ}) die away at the weak
magnetic field limit. Those that remain are the directional
dependence of the curvature contribution, together with the
curvature and shear Laplacians.

\subsubsection{Perturbations in the Magnetic Energy Density}
%%%%%%%%%%%%%%%%%%%%%%%%%%%%%%%%%%%%%%%%%%%%%%%%%%%%%%%%%%%%
For the linear evolution of the magnetic energy density gradients
we start from the definition of ${\cal B}$. Taking the time
derivative of Eq. (\ref{scalars}c)), we employ expression
(\ref{Pi}), the linear conservation law (\ref{medc}), the
background relations (\ref{sPi}) (guaranteing that
$\sigma_{ab}n^an^b={\textstyle{2\over3}}\lambda$ to zero-order)
and (\ref{bH2}), the linear result (\ref{bA}) and the propagation
equation (\ref{bdotDel}) to find
\begin{eqnarray}
\dot{{\cal B}}&=&\frac{4}{3(1+w)}\dot{\Delta}\nonumber\\&{}&-
\frac{4\Theta}{3(1+w)}\left\{\left[w-c_{\rm
s}^2(1-\zeta)-{\textstyle{4\over3}}\zeta\right]\Delta_{ab}n^an^b+
\left[w-c_{\rm s}^2\left(1-\zeta\left(1-c_{\rm
a}^2\right)\right)+{\textstyle{2\over3}}\zeta\right]
\Delta_{ab}v^{ab}\right\}
\nonumber\\&{}&-{\textstyle{4\over3}}\Theta
\left[\zeta\left(1+{\textstyle{1\over2}}c_{\rm a}^2\right){\cal
B}_{ab}n^an^b-{\textstyle{1\over2}}\zeta\left(1-c_{\rm
a}^2\right){\cal B}_{ab}v^{ab}\right]
\nonumber\\&{}&+{\textstyle{4\over3}}c_{\rm a}^2\zeta\Theta{\cal
K}_{ab}n^an^b
+2S^2\tilde{\nabla}^2\left(\sigma_{ab}n^an^b\right)\,,
\label{bdotcB}
\end{eqnarray}
where the shear Laplacian is the only effect of the anisotropic
background that remains at the weak magnetic field limit. In
deriving Eq. (\ref{bdotcB}) we have employed the linear
commutation law
\begin{equation}
\left(\tilde{\nabla}^2f\right)^{.}=\tilde{\nabla}^2\dot{f}-
{\textstyle{2\over3}}\Theta\tilde{\nabla}^2f-
2\sigma^{ab}\tilde{\nabla}_a\tilde{\nabla}_bf +\dot{f}A\,,
\label{tLcl}
\end{equation}
where $\tilde{\nabla}_af$ vanishes in the background. The above
follows from linearising the 3-divergence of (B18) in \cite{Du}.

\subsubsection{Perturbations in the Spatial Curvature}
%%%%%%%%%%%%%%%%%%%%%%%%%%%%%%%%%%%%%%%%%%%%%%%%%%%%%%
On using the barotropic expressions (\ref{bA}) and
(\ref{bardotsigma}), Eq. (\ref{dotcR}) is recast as
\begin{eqnarray}
\dot{{\cal R}}&=&-{\textstyle{2\over3}}\Theta
\left\{\left[1+2c_{\rm
a}^2\left(1+{\textstyle{1\over2}}\zeta\right)+2\zeta\right]{\cal
R}_{ab}n^an^b +(1-\zeta){\cal
R}_{ab}v^{ab}\right\}\nonumber\\&{}&+ \frac{4c_{\rm
s}^2\Theta}{3(1+w)S^2}\left[(1-\zeta)\Delta_{ab}n^an^b+\left(1-c_{\rm
a}^2\right)\left(1+{\textstyle{1\over2}}\zeta\right)\Delta_{ab}v^{ab}\right]
\nonumber\\&{}&+ \frac{2c_{\rm
a}^2\Theta}{3S^2}\left[(1-\zeta){\cal B}_{ab}n^an^b
+\left(1+{\textstyle{1\over2}}\zeta\right){\cal
B}_{ab}v^{ab}\right]\,,  \label{vdotcR}
\end{eqnarray}
which reduces to Eq. (40) in \cite{TB2} at the weak field limit.
For a relatively strong field with $c_{\rm a}^2$, $\zeta\sim1$,
the kinematic anisotropies (represented by $\zeta$) complicate the
overall magnetic impact on curvature. One faces a highly
anisotropic situation, where disturbances parallel to the
background field increase the average curvature, while
perturbations orthogonal to $H_a$ have the opposite effect and
vice versa. If we ignore all $\zeta$-contributions in  Eq.
(\ref{vdotcR}), the magneto-curvature effect on ${\cal R}$ is
imprinted in the direction dependent term $2c_{\rm a}^2{\cal
R}_{ab}n^an^b$. The latter boosts the smoothing effect of the
expansion on 3-curvature perturbations in agreement with
\cite{TB2}. In fact, the two effects coincide when ${\cal
R}_{ab}n^an^b={\textstyle{1\over3}}{\cal R}$.

It should be emphasised that all our equations refer to an
arbitrarily strong magnetic field. So far, the only restriction is
that the background field is an eigenvector of the shear tensor,
which translates into a simple algebraic dependence of the
zero-order shear and electric Weyl tensors on $\Pi_{ab}$ (see Eqs.
(\ref{sPi}) and (\ref{eWPi})). Next, we will apply these formulae
to the case of a relatively weak background magnetic field.

%%%%%%%%%%%%%%%%%%%%%%%%%%%%%%%%%%%%%%%
\section{The Weak Magnetic Field Limit}
%%%%%%%%%%%%%%%%%%%%%%%%%%%%%%%%%%%%%%%
The observed anisotropy of the CMB imposes strict limits on the
strength of any potential primordial magnetic field at the time of
decoupling. Also, Helium-4 measurements restrict the magnitude of
such field at the time of nucleosynthesis. All point towards a
relatively weak magnetic field, with current estimates arguing for
a field strength of $\sim10^{-8}G$ at present. Here, on the basis
of these observations, we assume a background magnetic field weak
compared to the dominant isotropically distributed matter
component (i.e. $h$, $c_{\rm a}^2\simeq h/(1+w)\ll1$). Given that
magnetism is the sole source of the background anisotropy, it is
reasonable to assume that the shear impact is also weak to zero
order (i.e. $\zeta\ll1$). At this limit, the unperturbed and
linear equations simplify considerably.

%%%%%%%%%%%%%%%%%%%%%%%%%%%%%%%%%
\subsection{Background Evolution}
%%%%%%%%%%%%%%%%%%%%%%%%%%%%%%%%%
Given that the source of the anisotropy is weak, one expects that
certain key figures of our Bianchi-I background will approach
their Friedmannian counterparts. Indeed, when $\zeta\ll1$ the
magnetic energy density (see Eq. (\ref{bH2})) evolves unaffected
by the background anisotropy
\begin{equation}
\left(H^2\right)^.= -{\textstyle{4\over3}}\Theta H^2\,.
\label{wmdc}
\end{equation}
The same is also true for the fluid energy density, given the
absence of magnetic terms in Eq. (\ref{bedc}).\footnote{For
radiation $w={\textstyle{1\over3}}$ and Eq. (\ref{doth}) implies
that $\dot{h}={\textstyle{4\over3}}\Theta\zeta h$. Also, the shear
impact parameter evolves as
$\dot{\zeta}=-{\textstyle{1\over3}}\Theta(\zeta+h)$, given that
$\zeta\equiv\lambda/\Theta$ and using Eqs. (\ref{dotl}),
(\ref{wbRay}), (\ref{wbFe}) and $\Lambda=0$. Thus, when there is
no magnetic field (i.e. $h=0$), the expansion anisotropy drops
rapidly. The field preserves the anisotropy so that $\zeta$ does
not tend to zero. Instead, $\dot{\zeta}\rightarrow0$ at late times
and $h$ experiences a slow logarithmic decay
\begin{equation}
h=\frac{h_0}{1+2\ln\left(t/t_0\right)}\,, \label{hlimit}
\end{equation}
in agreement with the `quasi-static' solution obtained in
\cite{Z}.}

The weakness of the field also means that the average volume
expansion of our Bianchi~I background approaches that of the
Friedmann model. When $h$, $\zeta\ll1$, the zero-order
Raychaudhuri equation (\ref{bRay}) gives
\begin{equation}
\dot{\Theta}=-{\textstyle{1\over3}}\Theta^2-
{\textstyle{1\over2}}\mu(1+3w)+\Lambda\,,  \label{wbRay}
\end{equation}
as in isotropic FRW cosmologies. The generalised Friedmann
equation (\ref{Fe2}) also acquires an FRW profile
\begin{equation}
\mu-{\textstyle{1\over3}}\Theta^2+\Lambda=0\,, \label{wbFe}
\end{equation}
given the spatial flatness of the Bianchi~I spacetime. This
guarantees that the dimensionless parameter $r$ in Eqs.
(\ref{bdotcZi}), (\ref{bdotcZ}) vanishes at the weak field limit.

It should be emphasised that these results do not, in any case,
suggest that the unperturbed Bianchi~I model has now been reduced
to a simple FRW cosmology. Our background universe is still
anisotropic, supplied with magnetic stresses, shear and an
electric Weyl tensor. However, the weakness of the magnetic field
means that, to leading order, the anisotropy no longer affects
certain key evolutionary aspects, such as the rate of the average
volume expansion.

%%%%%%%%%%%%%%%%%%%%%%%%%%%%%
\subsection{Linear Evolution}
%%%%%%%%%%%%%%%%%%%%%%%%%%%%%
\subsubsection{Kinematics}
%%%%%%%%%%%%%%%%%%%%%%%%%%
For a weak background magnetic field, Raychaudhuri's equation,
written in terms of the deceleration parameter, becomes
\begin{equation}
{\textstyle{1\over3}}\Theta^2{\rm q}=
{\textstyle{1\over2}}\mu(1+3w)+2\sigma^2+ \frac{c_{\rm
s}^2}{(1+w)S^2}\Delta+ \frac{c_{\rm a}^2}{2S^2}{\cal B}- c_{\rm
a}^2{\cal R}_{ab}n^an^b-\Lambda\,.  \label{wq}
\end{equation}
Relative to Eq. (\ref{q}), all the directional dependences in the
contributions from matter aggregations have been smoothed out. On
the other hand, comparing to Eq. (36) in \cite{TB2}, we notice the
extra shear term, which adds to the total deceleration. The
anisotropic background has also refined the curvature
contribution, which is now confined along the field lines only. In
agreement with the FRW treatment, the magneto-geometrical term
accelerates models with positive local curvature, but slows down
negatively curved regions. The effect observed here reduces
exactly to that identified by the FRW analysis when ${\cal
R}_{ab}n^an^b={\textstyle{1\over3}}{\cal R}$. Generally, however,
\begin{equation}
{\cal R}_{ab}n^an^b={\textstyle{1\over3}}{\cal R}-
{\textstyle{1\over3}}\mu h- {\textstyle{2\over3}}\sigma^2-
{\textstyle{1\over3}}\Theta\sigma_{ab}n^an^b+E_{ab}n^an^b\,.
\label{wcR1}
\end{equation}
Note that the impact of the $\sigma^2$-term in Eq. (\ref{wq})
becomes negligible provided that the weak field restriction is
imposed upon the actual universe, as well as the background.

Given the weakness of the background magnetic field, the vorticity
propagation equation reduces to
\begin{equation}
\dot{\omega}_a+
{\textstyle{2\over3}}\left(1-{\textstyle{3\over2}}c_{\rm
s}^2\right) \Theta\omega_a=
-\frac{1}{2\mu(1+w)}H^b\tilde{\nabla}_b{\rm curl}H_a+ \frac{c_{\rm
s}^2c_{\rm a}^2}{2\mu
h(1+w)S^2}\varepsilon_{abc}\Pi^{bd}\Sigma^c{}_d\,. \label{wdoto}
\end{equation}
For dust, the above agrees with the FRW equation (37) in
\cite{TB2}.

Assuming a weak background magnetic field has very little effect
on the shear evolution. In fact, for $c_{\rm a}^2$, $\zeta\ll1$
Eq. (\ref{bardots}) gives
\begin{eqnarray}
\dot{\sigma}_{\langle ab\rangle}&=&-
{\textstyle{2\over3}}\Theta\sigma_{ab}- \frac{c_{\rm
s}^2}{(1+w)S^2}\Sigma_{ab}- \frac{c_{\rm a}^2}{2S^2}{\cal
B}_{\langle ab\rangle}+ {\textstyle{1\over3}}c_{\rm a}^2{\cal
R}_{\langle ab\rangle}+
{\textstyle{1\over2}}\Pi_{ab}\nonumber\\&{}& + \frac{c_{\rm
a}^2}{\mu hS}H^c\tilde{\nabla}_c{\cal M}_{\langle ab\rangle}-
E_{ab}-\sigma_{c\langle a}\sigma^c{}_{b\rangle}+ \frac{c_{\rm
s}^2c_{\rm a}^2}{\mu h (1+w)S^2} \Pi_{c\langle
a}\Delta^c{}_{b\rangle}\nonumber\\&{}&- \frac{c_{\rm a}^2}{\mu
h}\Pi_{cd} {\cal R}_{\langle
a\hspace{1mm}b\rangle}^{\hspace{2mm}c\hspace{2mm}d}- \frac{4c_{\rm
a}^2\Theta}{3\mu h}\Pi^c{}_{\langle a}\varepsilon_{b\rangle
cd}\omega^d\,, \label{wdots}
\end{eqnarray}
which is considerably more complicated than Eq. (38) in
\cite{TB2}. This is not surprising, given that the shear reflects
the intrinsic spatial anisotropy of the Bianchi~I spacetime.

\subsubsection{Dynamics}
%%%%%%%%%%%%%%%%%%%%%%%%
At the weak field limit, Eq. (\ref{bdotDel}) that monitors
gradients in the fluid energy density, reduces to\footnote{Looking
at Eq. (\ref{bdotDel}), it becomes clear that the
$\zeta$-contributions along and transverse to the field lines are
relatively weak when $w$, $c_{\rm s}^2\neq0$. Even in the dust
era, when $w=0=c_{\rm s}^2$, the effect of these terms is
negligible. Indeed, on using definition (\ref{Delij}) one can
verify that $\dot{\Delta}\pm\zeta\Theta\Delta\simeq\dot{\Delta}$
for $\zeta\ll1$. Similarly, one shows that $\dot{\Delta}\pm c_{\rm
a}^2\Theta\Delta\simeq\dot{\Delta}$ when $c_{\rm a}^2\ll1$.}
\begin{equation}
\dot{\Delta}=w\Theta\Delta-(1+w){\cal Z}+
{\textstyle{1\over2}}(1+w)c_{\rm a}^2\Theta{\cal B}-(1+w)c_{\rm
a}^2 \Theta {\cal K}_{ab}n^an^b\,, \label{wdotDel}
\end{equation}
Again, all directional dependences in the contribution of $\Delta$
have died away. The difference between the above and its FRW
counterpart, given by Eq. (54) in \cite{TB2}, is in their
magneto-curvature terms. The two equations coincide when ${\cal
K}_{ab}n^an^b={\textstyle{1\over3}}{\cal K}$.

When the field is relatively weak, $r=0$ and Eq. (\ref{bdotcZ}),
for the propagation of the expansion gradients simplifies to
\begin{eqnarray} \dot{{\cal Z}}&=&
-{\textstyle{2\over3}}\Theta{\cal Z}-
{\textstyle{1\over2}}\mu\Delta+ {\textstyle{1\over4}}(1+w)\mu
c_{\rm a}^2{\cal B}- {\textstyle{3\over2}}(1+w)\mu c_{\rm
a}^2{\cal K}_{ab}n^an^b \nonumber\\&{}&- \frac{c_{\rm
s}^2}{1+w}\widetilde{\nabla}^2\Delta- {\textstyle{1\over2}}c_{\rm
a}^2\widetilde{\nabla}^2{\cal B}+ c_{\rm
a}^2\widetilde{\nabla}^2\left({\cal K}_{ab}n^an^b\right)-
2\zeta\Theta
S^2\widetilde{\nabla}^2\left(\sigma_{ab}n^an^b\right)\,,
\label{wdotcZ}
\end{eqnarray}
since all the directional differences in the ${\cal Z}$, ${\cal
D}$ and ${\cal B}$ contributions have vanished. Nevertheless, the
anisotropy of the Bianchi~I background is still imprinted in Eq.
(\ref{wdotcZ}) via ${\cal K}_{ab}n^an^b$ and the new shear term,
both of which are direction dependent. Note that the Laplacian
contributions become negligible on wavelengths larger than the
Hubble horizon. Also, when ${\cal K}_{ab}n^an^b$ takes the average
value ${\textstyle{1\over3}}{\cal K}$ all the differences in the
spatial curvature terms, between Eq. (\ref{wdotcZ}) above and  Eq.
(55) in \cite{TB2}, disappear.

At the weak field limit, almost all the directional effects in Eq
(\ref{bdotcB}) are removed. The magnetic energy gradients
propagate as
\begin{equation} \dot{{\cal B}}=\frac{4}{3(1+w)}\dot{\Delta}+
\frac{4\left(c_{\rm s}^2-w\right)\Theta}{3(1+w)}\Delta+
2S^2\widetilde{\nabla}^2\left(\sigma_{ab}n^an^b\right)\,,
\label{wdotcB}
\end{equation}
where terms quadratic in the smallness parameters $c_{\rm a}^2$
and $\zeta$ have been left out. Relative to Eq. (61) in
\cite{TB2}, there is an extra input depending on the shear
component parallel to $H_a$. This effect, which results from the
background anisotropy, becomes negligible on superhorizon scales.

Finally, a weak background magnetic field means that the all
anisotropic effects in Eq. (\ref{vdotcR}) are smoothed out. The
latter reduces to its FRW counterpart
\begin{equation}
\dot{{\cal K}}=\frac{4c_{\rm s}^2\Theta}{3(1+w)}\Delta+
{\textstyle{2\over3}}c_{\rm a}^2\Theta{\cal B}\,, \label{wdotcK}
\end{equation}
where 3-curvature perturbations are now represented by the
dimensionless scalar ${\cal K}\equiv S^2{\cal R}$ (compare to Eq.
(57) in \cite{TB2}).

%%%%%%%%%%%%%%%%%%%%%%%%%%%%%%
\section{Particular Solutions}
%%%%%%%%%%%%%%%%%%%%%%%%%%%%%%
\subsection{Large scales}
%%%%%%%%%%%%%%%%%%%%%%%%%
On wavelengths larger that the Hubble horizon the Laplacian terms
in Eqs. (\ref{wdotcZ}) and (\ref{wdotcB}) are negligible. On these
scales, the evolution of magnetised density gradients is monitored
by the following system of ordinary differential equations
\begin{eqnarray}
\dot{\Delta}&=&w\Theta\Delta-(1+w){\cal Z}+
{\textstyle{1\over2}}(1+w)c_{\rm a}^2\Theta{\cal B}- (1+w)c_{\rm
a}^2 \Theta {\cal K}_{ab}n^an^b\,, \label{lwdotDel}\\ \nonumber\\
\dot{{\cal Z}}&=& -{\textstyle{2\over3}}\Theta{\cal Z}-
{\textstyle{1\over2}}\mu\Delta+ {\textstyle{1\over4}}(1+w)\mu
c_{\rm a}^2{\cal B}- {\textstyle{3\over2}}(1+w)\mu c_{\rm
a}^2{\cal K}_{ab}n^an^b\,, \label{lwdotcZ}\\ \dot{{\cal
B}}&=&\frac{4}{3(1+w)}\dot{\Delta}+ \frac{4\left(c_{\rm
s}^2-w\right)\Theta}{3(1+w)}\Delta\,, \label{lwdotcB}\\
 \dot{{\cal K}}&=&\frac{4c_{\rm
s}^2\Theta}{3(1+w)}\Delta+ {\textstyle{2\over3}}c_{\rm
a}^2\Theta{\cal B}\,. \label{lwdotcK}
\end{eqnarray}
When ${\cal K}_{ab}n^an^b={\textstyle{1\over3}}{\cal K}$ the above
coincide with the equations obtained by the FRW-based analysis of
\cite{TB2}.\footnote{Any numerical differences between our
equations and solutions and those of \cite{TB2,TM} are due to
differences in the Alfv\'en speed definitions.} In this occasion,
which one might call the case of `average' or `isotropic'
curvature perturbations, the predictions of the Bianchi~I analysis
are identical to those of the FRW-based treatment.

Given that the above system closes whenever ${\cal
K}_{ab}n^an^b\propto{\cal K}$, it is worth considering two
additional critical cases. The `minimum' curvature case, when
${\cal K}_{ab}n^an^b=0$, and that of `maximum' curvature with
${\cal K}_{ab}n^an^b={\cal K}$ (i.e. for ${\cal K}_{ab}v^{ab}=0$).
In the former situation, the equations are clear of any
magneto-curvature complexities and the evolution of $\Delta$
proceeds as predicted by the FRW treatment. In the case of maximum
curvature, however, there is a small numerical difference between
the Bianchi~I equations and those of the almost-FRW analysis. Note
that the isotropic and the zero curvature cases have been already
addressed in \cite{TB2} and subsequently refined in \cite{TM}.
Here, we will reproduce these results for comparison reasons and
concentrate on the case of maximum curvature. Despite the fact
that all three cases are rather special, they are quite important.
As we shall see next, their study illustrates the effect of the
magneto-curvature coupling on the growth of linear density
perturbations.

\subsubsection{Radiation Era}
%%%%%%%%%%%%%%%%%%%%%%%%%%%%%
During the radiation era $w=c_{\rm s}^2={\textstyle{1\over3}}$,
the fluid density falls as $\mu\propto S^{-4}$ and the Alfv\'{e}n
speed can be treated as a constant (i.e. $\left(c_{\rm
a}^2\right)^{.}=0$).\footnote{When radiation dominates the
evolution of the background Alfv\'{e}n speed depends on $\zeta$,
the shear impact parameter (see Eq. (\ref{dotAlf})). At the weak
field limit this dependence translates into a non-linear
logarithmic decay for $c_{\rm a}^2$, identical to that of $h$ (see
Eq. (\ref{hlimit})). In any case, the coupling between $c_{\rm
a}^2$ and $\mu$ or $\Theta$ in Eqs. (\ref{lwdotDel}),
(\ref{lwdotcZ}) and (\ref{lwdotcK}), ensures that the Alfv\'{e}n
speed behaves as if it was time independent.} Expressed in terms
of the scale factor, to facilitate the solution, the system
(\ref{lwdotDel})-(\ref{lwdotcK}) writes
\begin{equation}
S^2\frac{d^2\Delta^{(\nu)}}{dS^2}=2\Delta^{(\nu)}-
{\textstyle{4\over3}}c_{\rm a}^2{\cal B}^{(\nu)}+ 8c_{\rm
a}^2{\cal K}^{(\nu)}\,, \label{rlwdotDel3}
\end{equation}
\begin{equation}
\frac{d{\cal B}^{(\nu)}}{dS}=\frac{d\Delta^{(\nu)}}{dS}\,,
\label{rlwdotcB3}
\end{equation}
\begin{equation}
\\ \mbox{}\nonumber\\ S\frac{{\cal
K}^{(\nu)}}{dS}=\Delta^{(\nu)}+ 2c_{\rm a}^2{\cal B}^{(\nu)}\,.
\label{rlwdotcK3}
\end{equation}
where the expansion gradients have been decoupled from the
equations and we have set $\Lambda=0$. Also, $\Delta^{(\nu)}$ is
the $\nu$-th harmonic component of $\Delta$ and similar
decompositions have been applied to the rest of the inhomogeneity
scalars. The above has the following power law solution
\begin{equation}
\Delta^{(\nu)}(S)=\Delta^{(\nu)}_0+
\sum_{\alpha}\Delta^{(\nu)}_{\alpha}S^{\alpha}\,, \label{rlwDel}
\end{equation}
where $\Delta^{(\nu)}_0$, $\Delta^{(\nu)}_{\alpha}$ are arbitrary
positive constants and the parameter $\alpha$ satisfies the cubic
equation
\begin{equation}
\alpha^3-\alpha^2-2\left(1-{\textstyle{2\over3}}c_{\rm
a}^2\right)\alpha-8c_{\rm a}^2=0\,, \label{z}
\end{equation}
to lowest order in $c_{\rm a}^2$. This has one positive and two
negative roots that correspond to one growing and two decaying
modes. For small $c_{\rm a}^2$, the roots are perturbatively found
to be
\begin{equation}
\alpha=\left\{\begin{array}{l} -1+{\textstyle{28\over9}}c_{\rm
a}^2+{\cal O}(c_{\rm a}^4)\,,\\\\2+{\textstyle{8\over9}}c_{\rm
a}^2+{\cal O}(c_{\rm a}^4)\,,\\\\  -4c_{\rm a}^2+{\cal O}(c_{\rm
a}^4)\,,
\end{array}\right.  \label{mz123}
\end{equation}
which implies the following evolution
\begin{equation}
\Delta^{(\nu)}=\Delta^{(\nu)}_0+\Delta^{(\nu)}_+
S^{2+{\textstyle{8\over9}}c_{\rm a}^2}+\Delta^{(\nu)}_{1-}
S^{-1+{\textstyle{28\over9}}c_{\rm a}^2}+\Delta^{(\nu)}_{2-}
S^{-4c_{\rm a}^2} \label{rlwDel1}
\end{equation}
for the density contrast. In the absence of the magnetic field,
namely for $c_{\rm a}^2=0$, the above reduces to the standard
solution of cosmological density perturbations (see \cite{P} for
example). Hence, the field presence has led to a slight increase
of the growing mode. As we shall see next, this increase results
from the coupling between magnetism and geometry. In particular,
for minimum curvature perturbations, ${\cal K}_{ab}n^an^b=0$ and
\begin{equation}
\alpha=\left\{\begin{array}{l} -1+{\textstyle{4\over9}}c_{\rm
a}^2+{\cal O}(c_{\rm a}^4)\,,\\\\ 2-{\textstyle{4\over9}}c_{\rm
a}^2+{\cal O}(c_{\rm a}^4)\,,
\end{array}\right.  \label{0z12}
\end{equation}
to lowest order in $c_{\rm a}^2$ \cite{TB2}. Thus, in the absence
of any curvature contributions the pure magnetic effect is to
reduce the growth rate of the density contrast. After
recombination, an analogous reduction of the growing mode was also
identified both by the relativistic analysis (see \cite{TB2}) and
by the Newtonian treatment (see \cite{RR}). In \cite{TM}, this
particular effect was attributed to the `frozen in' property of
the field. The latter combines with the dilution of the magnetic
force-lines caused by the expansion to oppose gravitational
collapse. Alternatively, for isotropic curvature perturbations
${\cal K}_{ab}n^an^b={\textstyle{1\over3}}{\cal R}$ and
\begin{equation}
\alpha=\left\{\begin{array}{l} -1+{\textstyle{4\over3}}c_{\rm
a}^2+{\cal O}(c_{\rm a}^4)\,,\\\\2+{\cal O}(c_{\rm a}^4)\,,\\\\
-{\textstyle{4\over3}}c_{\rm a}^2+{\cal O}(c_{\rm a}^4)\,,
\end{array}\right.  \label{iz123}
\end{equation}
to lowest order in $c_{\rm a}^2$ \cite{TM}. In this occasion the
density contrast grows exactly as in non-magnetised cosmologies.

Comparing results (\ref{mz123}), (\ref{0z12}) and (\ref{iz123}) we
notice that as the magneto-curvature contribution gets stronger,
from ${\cal K}_{ab}n^an^b=0$ to ${\cal
K}_{ab}n^an^b={\textstyle{1\over3}}{\cal K}$ and finally to ${\cal
K}_{ab}n^an^b={\cal K}$, the growing mode of $\Delta$ successively
increases. In fact, {\em by the time the curvature input has taken
its maximum value, the pure magnetic effect has been completely
reversed}. In this case, the density contrast grows faster than
its counterpart in non-magnetised cosmologies. Again, the reason
behind this effect is the special form of the anisotropic magnetic
stress tensor. The latter ensures that there is a negative
pressure, a tension, along the direction the field lines. As
mentioned in Sec. 5, when combined with geometry, this property of
the field opposes the action of the magnetic energy density
gradients (see Eqs. (\ref{lwdotDel}), (\ref{rlwdotDel3})).

Undoubtedly, the weakness of the background field means that the
aforementioned magnetic effects are small `corrections' to the
standard evolution of density perturbations. Interestingly,
however, whether the field boosts or inhibits gravitational
collapse may entirely depend on the coupling between magnetism and
geometry.

%%%%%%%%%%%%%%%%%%%%%%%%%%%%%%%%%%%%%%
\subsection{Dust Era and Small Scales}
%%%%%%%%%%%%%%%%%%%%%%%%%%%%%%%%%%%%%%
In the dust era Eqs. (\ref{lwdotDel}), (\ref{lwdotcZ}) and
(\ref{lwdotcK}) guarantee that, to lowest order in $c_{\rm a}^2$,
3-curvature has no effect on the propagation of $\Delta$. In other
words, after recombination long-wavelength magnetised density
inhomogeneities proceed exactly as predicted by the Friedmannian
treatment \cite{TB2,TM}. Thus, it becomes clear that the FRW
framework offers a quite accurate treatment of large-scale
magnetised cosmological perturbations in almost all situations.

For a general ${\cal K}_{ab}n^an^b$, however, equations
(\ref{wdotDel})-(\ref{wdotcK}) are not adequate to describe the
evolution of $\Delta$. In order to close the system, one requires
the propagation equation for ${\cal K}_{ab}n^an^b$ instead of
(\ref{wdotcK}). This, in turn, is immediately coupled to the shear
and the electric Weyl tensors via the first order relation
\begin{equation}
{\cal K}_{ab}n^an^b={\textstyle{1\over3}}{\cal K}-
{\textstyle{1\over3}}S^2\mu h- {\textstyle{2\over3}}S^2\sigma^2-
{\textstyle{1\over3}}\Theta
S^2\sigma_{ab}n^an^b+S^2E_{ab}n^an^b\,, \label{cK1}
\end{equation}
obtained form Eq. (\ref{wcR1}). It follows that when the
anisotropy in the curvature perturbation is `non-critical' the
problem becomes very complicated to address analytically. In this
case, as well as on subhorizon scales, one has to resort to
numerical methods for solutions. Alternatively, one might employ
dynamical system techniques (see \cite{WE} for an overview) to
obtain qualitative answers. This approach has been employed by
\cite{Lb} to analyse an exact Bianci~I cosmological model
permeated by source-free large-scale magnetic field. Dynamical
system methods were also used in \cite{Du} to study qualitatively
non-magnetised density inhomogeneities within perturbed Bianchi~I
cosmologies. Note that one may invoke fluid backreaction arguments
and ignore, to first order, the small scale shear effects. In this
case the system (\ref{wdotDel})-(\ref{wdotcK}) reduces to its FRW
counterpart on every scale provided ${\cal
K}_{ab}n^an^b={\textstyle{1\over3}}{\cal K}$ \cite{TB2}.

%%%%%%%%%%%%%%%%%%%%
\section{Conclusion}
%%%%%%%%%%%%%%%%%%%%
We have pursued the study of magnetised cosmological perturbations
within a perturbed Bianchi~I spacetime, in the first attempt of
this nature. So far, the subject has been only addressed within
the limits perturbed FRW cosmologies, given the observed high
isotropy of the CMB spectrum. Strictly mathematically speaking,
however, a spatially isotropic spacetime cannot naturally
accommodate large-scale magnetic fields. Even when the background
field is is treated as random or weak, there are problems related
to the generic anisotropy of the magnetic stresses. Although it is
physically plausible that weak fields can be adequately treated
within the almost-FRW limit, mathematically speaking the Friedmann
spacetimes are unsuitable hosts for cosmic magnetism. In this
respect, the spatially anisotropic Bianchi~I background provides a
better framework for studying the magnetic effects on cosmological
perturbations. Moreover, such study can be easily applied to
primordial magnetic fields of arbitrary strengths, if required.

The search for mathematical consistency was not the only
motivation behind this work. The relativistic, but almost-FRW,
treatments of \cite{TB2,TM} suggested a number of unexpected
magnetic effects on the kinematics and the dynamics of the
universe. All of them seemed to derive from an intriguing coupling
between magnetism and geometry. The potential importance of these
magneto-curvature effects meant that a new study, this time in a
more natural environment, was necessary. The Bianchi~I model,
which has long been known to be the simplest anisotropic spacetime
able to accommodate cosmological magnetic fields, provided just
this environment.

We began by assuming that the entire background anisotropy was due
to the magnetic presence, in order to isolate the anisotropic
effects of the field. Thus, in the background kinematic
anisotropies as well as anisotropies in the spacetime geometry are
both the result of the magnetic stresses. In mathematical terms
this is achieved by demanding that, to zero order, the magnetic
vector is an eigenvector of the shear tensor. Although this scheme
may not represent the most general case, it is sufficient for
studying the lowest order effects of the magnetic stresses at the
onset of structure formation. Any zero-order anisotropies that are
not the result of the magnetic presence are irrelevant to our
study and therefore ignored. At the linear level, however, the
field is no longer assumed to be a shear eigenvector. Thus, to
first order the shear does not entirely depend on the magnetic
stresses. Our first step was to linearise the non-linear formulae
of \cite{TB1} about a Bianchi~I spacetime, and also provide a
couple of new but necessary first-order relations. As in
\cite{TB2,TM}, we applied our equations to the case of a
barotropic medium and then examined in detail the
magnetohydrodynamical effects on the kinematics and the dynamics
of the universe. It was at this level that we were first able to
identify the implications of the anisotropic background.

In particular, it became clear that some of the effects of the
FRW-based treatment were in fact direction dependent. To be
precise, a number of quantities were found to contribute
differently parallel to the field lines than normally to them. For
example, for a relatively strong magnetic field, the decelerating
effect from matter aggregations was no longer isotropic. We found
that only density perturbations parallel to the direction of the
zero-order field vector can actually affect the average volume
expansion. In contrast, density gradients perpendicular to the
magnetic force-lines get weaker as the strength of the field
increases. Nevertheless, the isotropy of this effect was
immediately reinstated at the weak field limit to match the
predictions of \cite{TB2}.

The role of the background field stresses was further illustrated
through the magneto-curvature contribution to the deceleration
parameter. In \cite{TB2}, the isotropy of the unperturbed FRW
spacetime meant that the curvature effects were direction
independent. Here, however, we found that their contribution is
entirely confined along the zero-order field vector. As in
\cite{TB2}, the resulting magneto-curvature term adds to the
deceleration of disturbances with local negative spatial
curvature, but accelerates positively curved regions. This seemed
odd at first, given that positive curvature is traditionally
associated with gravitational collapse. The explanation came from
the directional dependence of the curvature effect. Recall that
there is a negative pressure, a tension, in the direction of the
magnetic lines of force. This is another way of saying that the
field lines are elastic, opposing any attempt to distort their
equilibrium pattern. When coupled to geometry, this magnetic
property ensures that the field will react to any changes in the
spatial curvature by modifying the local expansion rate
accordingly. This relativistic effect is closely analogous to the
classical curvature stress exerted by field lines with a nonzero
curvature radius. Note that, qualitatively speaking, the
aforementioned effect does not depend on the field's relative
strength.

Similarly, the role of the field as a source of universal rotation
is also independent of its strength. As in \cite{TB2}, we find
that when the curl of the magnetic vector varies along the
direction of the background filed, cosmic magnetism generates
vortices in the fluid flow. However, the magnetic effect on the
evolution of these vortices differs according to the strength of
the field. In particular, the presence of strong magnetic fields
can modify the dilution rate of such rotational disturbances.

Our study also facilitated a direct test for the FRW-based
treatments of weakly magnetised density perturbations. We can see
now the number and the nature of the corrections due to the
background anisotropy, and identify the domains they mainly
affect. Therefore, we are in a better position to judge the
accuracy of the Friedmannian approximation. For an arbitrarily
strong magnetic field, we find several directional contributions
in the propagation equations of the basic inhomogeneity variables.
As before, the key direction is determined by the background
magnetic field. At this level, there is a considerable gap
separating the Bianchi~I equations from those of the FRW
approximation. At the weak field limit, however, these differences
are reduced dramatically. In particular, we find that only
curvature and shear effects propagate along the preferred
direction of the background field vector. Moreover, after
recombination, density perturbations proceed essentially
unaffected by curvature complications. Also, the shear
contributions, which result from the kinematical implications of
the background field, are confined to sub-horizon scales only. As
a result, the large-scale equations of the FRW analysis are
identical to those of the Bianchi~I treatment when ${\cal
R}_{ab}n^an^b$, the curvature perturbation along the field lines,
takes the `isotropic' value ${\textstyle{1\over3}}{\cal R}$. Also,
the two approaches are essentially indistinguishable in two
additional critical cases, when ${\cal R}_{ab}n^an^b=0$ and for
${\cal R}_{ab}n^an^b={\cal R}$. In the former case the curvature
perturbation parallel to $H_a$ is said to be `minimum' and in the
latter `maximum'. Conclusively, we feel justified to argue that,
at least on super-horizon scales, density perturbations are
accurately treated by the FRW approximation. Even on small scales,
one may be able to invoke backreaction arguments and ignore the
aforementioned shear effects from the linear perturbation
equations.

With regard to density perturbations, the Bianchi~I analysis has
lead to an additional, this time less trivial, result. In
particular, let us recall the results of the isotropic and the
minimum curvature cases mentioned above, which were addressed in
\cite{TB2,TM}. There, it was found that in the absence of any
curvature input (i.e. for ${\cal R}_{ab}n^an^b=0$) the magnetic
presence suppresses the growth rate of matter aggregations. In
contrast, when isotropic curvature contributions were allowed
(i.e. ${\cal R}_{ab}n^an^b={\textstyle{1\over3}}{\cal R}$), the
overall magnetic effect was reduced and the density gradients grew
as fast as in non-magnetised universes. This pattern was also
confirmed here. Moreover, for maximum curvature contributions,
that is when ${\cal R}_{ab}n^an^b$ increases from
${\textstyle{1\over3}}{\cal R}$ to ${\cal R}$, the density
contrast grows even faster than in magnetic-free cosmologies.
Thus, the coupling of the field to the geometry can actually
reverse the original magnetic impact on gravitational collapse.
The reason behind this effect is again the tension properties of
the magnetic force-lines, which leads to a curvature stress
proportional to $c_{\rm a}^2{\cal R}_{ab}n^an^b$. The latter tends
to counterbalance the action of the magnetic pressure gradients,
thus reversing the pure magnetic impact on linear density
condensations. As a result, in the magnetic presence, the density
contrast may even grow faster than in non-magnetised cosmologies.

%%%%%%%%%%%%%%%%%%%%%%%%%%
\section*{Acknowledgments}
%%%%%%%%%%%%%%%%%%%%%%%%%%
CGT was supported by PPARC. The authors would like to thank John
Wainwright, John Barrow, Jai-chan Hwang and Marco Bruni for
helpful comments.

%%%%%%%%%%%%%%%%%%
\section{Appendix}
%%%%%%%%%%%%%%%%%%
\subsection{Interpreting {\bf ${\cal D}_a$}}
%%%%%%%%%%%%%%%%%%%%%%%%%%%%%%%%%%%%%%%%%%%%
Following \cite{E2}, the connecting vector $\delta x^a$ between
two neighbouring worldlines (i.e. fundamental observers) satisfies
the restriction
\begin{equation}
u^b\nabla_b\delta x_a=\delta x^b\nabla_bu_a\,.  \label{cv}
\end{equation}
Also, the relative position vector, which connects the same two
points $P$, $P'$ on these worldlines at all times, is
\begin{equation}
\delta x_{\langle a\rangle}=h_{ab}\delta x^b\,, \label{rpv}
\end{equation}
and propagates as
\begin{equation}
\left(\delta x_{\langle a\rangle}\right)^{.}=
\left(\sigma_{ab}+\varepsilon_{abc}\omega^c+
{\textstyle{1\over3}}\Theta h_{ab}\right)\delta x^{\langle
b\rangle}\,. \label{rpvprop}
\end{equation}

Suppose now that $\mu(x_c)$ and $\mu(x_c+\delta x_{\langle
c\rangle})$ is the fluid density at $P$ and $P'$ respectively,
relative to a frame at $P$. Then, to first order,
\begin{eqnarray}
\mu(x_c+\delta x_{\langle c\rangle})&=&\mu(x_c)+
\left(\tilde{\nabla}^a\mu\right)\delta x_{\langle a\rangle}
\Rightarrow \nonumber\\ \delta\equiv\frac{\delta\mu}{\mu}&=&{\cal
X}^a \delta x_{\langle a\rangle}\,, \label{dmu/mu}
\end{eqnarray}
where $\delta\mu\equiv\mu(x_c+\delta x_{\langle
c\rangle})-\mu(x_c)$ and ${\cal
X}_a\equiv\tilde{\nabla}_a\mu/\mu$. Equation (\ref{rpvprop})
ensures that within a FRW spacetime
\begin{equation}
\delta x_{\langle a\rangle}=S\left(\delta x_{\langle
a\rangle}\right)_0\,, \label{rpvev}
\end{equation}
where $\left(\delta x_{\langle a\rangle}\right)_0$ is a constant.
Therefore, result (\ref{dmu/mu}) becomes
\begin{equation}
\delta=\left(\delta x^{\langle a\rangle}\right)_0{\cal D}_a\,,
\label{Dmu/mu}
\end{equation}
where ${\cal D}_a\equiv S{\cal X}_a$. Thus, in an almost-FRW
universe the comoving fractional density gradient ${\cal D}_a$
describes the spatial density variations between two neighbouring
fundamental observers. However, in a Bianchi~I spacetime the
background shear is non-zero and $\delta x_{\langle i\rangle}$ no
longer evolves according to (\ref{rpvev}). Consequently, in a
perturbed Bianchi~I universe ${\cal D}_a$ no longer describes the
spatial density changes seen by a pair of neighbouring fundamental
observers. Obviously, in the case of weak background shear
$\sigma/\Theta\ll1$ and the aforementioned complication is
negligible, and $D_a$ regains the desired interpretation.

\end{document}